\documentclass[floatfix,aps,unsortedaddress]{revtex4-1}
\usepackage{graphicx}
\usepackage{dcolumn}
\usepackage{bm}
\usepackage[utf8]{inputenc}
\usepackage{amssymb}
\usepackage{placeins}
\usepackage{amsmath}
\usepackage{amsthm}
\usepackage{commath}
\usepackage{subcaption}
\usepackage[margin=1in]{geometry}
\usepackage[citecolor=blue,colorlinks=true,linkcolor=blue]{hyperref}
\newcommand{\partder}[2]{\dfrac{\partial  #1}{\partial  #2}} 
\newcommand{\der}[2]{\dfrac{d #1}{d  #2}}

\begin{document}

\title{An adjoint method for gradient-based optimization of stellarator coil shapes}
\author{E. J. Paul}
\affiliation{Department of Physics, University of Maryland, College Park, MD 20742, USA}
\email{ejpaul@umd.edu}

\author{M. Landreman}
\affiliation{Institute for Research in Electronics and Applied Physics, University of Maryland, College Park, MD 20742, USA}

\author{A. Bader}
\affiliation{Department of Engineering Physics, University of Wisconsin, Madison, WI 53706, USA}

\author{W. Dorland}
\affiliation{Department of Physics, University of Maryland, College Park, MD 20742, USA}

\begin{abstract}
We present a method for stellarator coil design via gradient-based optimization of the coil-winding surface. The REGCOIL (Landreman 2017 \textit{Nucl. Fusion} \textbf{57} 046003) approach is used to obtain the coil shapes on the winding surface using a continuous current potential. We apply the adjoint method to calculate derivatives of the objective function, allowing for efficient computation of analytic gradients while eliminating the numerical noise of approximate derivatives. We are able to improve engineering properties of the coils by targeting the root-mean-squared current density in the objective function. We obtain winding surfaces for W7-X and HSX which simultaneously decrease the normal magnetic field on the plasma surface and increase the surface-averaged distance between the coils and the plasma in comparison with the actual winding surfaces. The coils computed on the optimized surfaces feature a smaller toroidal extent and curvature and increased inter-coil spacing. A technique for visualization of the sensitivity of figures of merit to normal surface displacement of the winding surface is presented, with potential applications for understanding engineering tolerances. 
\end{abstract}

\maketitle

\section{Introduction}
Stellarators confine particles by generating rotational transform with external coils. The 3-dimensional nature of a stellarator presents great opportunity, allowing a large space within which to find optimal plasma configurations. However, designing coils to produce the necessary non-axisymmetric magnetic field is a significant challenge for the stellarator program. The design of simple coils which can be reasonably engineered and produce a plasma with optimal physics properties is required in order for the steady-state, disruption-free confinement of optimized stellarators to be realized. 

Stellarator coils are usually designed to produce a target outer plasma boundary. The plasma boundary is separately optimized for various physics quantities, including magnetohydrodynamic (MHD) stability, neoclassical confinement, and profiles of rotational transform and pressure \cite{Nuhrenberg1988}. The coil shapes are then optimized such that one of the magnetic surfaces approximately matches the desired plasma surface. In general the desired plasma configuration can not be produced exactly due to engineering constraints on the coil complexity.

In addition to minimization of the magnetic field error, there are several factors that should be considered in the design of coils shapes. The winding surface upon which the currents lie should be sufficiently separated from the plasma surface to allow for neutron shielding to protect the coils, the vacuum vessel, and a divertor system. In a reactor, the coil-plasma distance is closely tied to the tritium breeding ratio and overall cost of electricity as it determines the allowable blanket thickness. The coil-plasma distance was targeted in the ARIES-CS study to reduce machine size \cite{Guebaly2008}. In practice the minimum feasible coil-plasma separation is a function of the desired plasma shape. Concave regions (such as the bean W7-X cross section) are especially difficult to produce \cite{Landreman2016} and require the winding surface to be near to the plasma surface. While decreasing the inter-coil spacing minimizes ripple fields, increasing coil-coil spacing allows adequate space for removal of blanket modules, heat transport plumbing, diagnostics, and support structures. The curvature of a coil should be below a certain threshold to allow for the finite thickness of the conducting material and to avoid prohibitively high manufacturing costs. The length of each coil should also be considered, as expense will grow with the amount of conducting material that needs to be produced. For these reasons, identifying coils with suitable engineering properties can impact the size and cost of a stellarator device. 

Most coil design codes have assumed the coils to lie on a closed toroidal winding surface enclosing the desired plasma surface. In \texttt{NESCOIL} \cite{Merkel1987}, the currents on this surface are determined by minimizing the integral-squared normal magnetic field on the target plasma surface. Using a stream function approach, the current potential on the winding surface is decomposed in Fourier harmonics. This takes the form of a least-squares problem which can be solved with a single linear system. The coil filament shapes can be obtained from the contours of the current potential. Because it is guaranteed to find a global minimum, \texttt{NESCOIL} is often used in the preliminary stages of the design process \cite{Spong2010, Ku2011, Drevlak2013}. It was used for the initial coil configuration studies for NCSX \cite{Pomphrey2001}. The W7-X coils were designed using an extension of \texttt{NESCOIL} which modified the winding surface geometry for quality of magnetic surfaces and engineering properties of the coils \cite{Beidler1990}. However, the inversion of the Biot-Savart integral by \texttt{NESCOIL} is fundamentally ill-posed, resulting in solutions with amplified noise. The \texttt{REGCOIL} \cite{Landreman2017} approach addresses this problem with Tikhonov regularization. Here the surface-average-squared current density, corresponding to the squared-inverse distance between coils, is added to the objective function. With the addition of this regularization term, \texttt{REGCOIL} is able to simultaneously increase the minimum coil-coil distances and improve reconstruction of the desired plasma surface over \texttt{NESCOIL} solutions. In this work we build on the \texttt{REGCOIL} method to optimize the current distribution in 3 dimensions. The current distribution on a single winding surface is computed with \texttt{REGCOIL}, and the winding surface geometry is optimized to reproduce the plasma surface with fidelity and improve engineering properties of the coil shapes. 

Other nonlinear coil optimization tools exist which evolve discrete coil shapes rather than continuous surface current distributions. Drevlak's \texttt{ONSET} code \cite{Drevlak1998} optimizes coils within limiting inner and outer coil surfaces. The \texttt{COILOPT} \cite{Strickler2002,Strickler2004} code, developed for the design of the NCSX coil set \cite{Zarnstorff2001}, optimizes coil filaments on a winding surface which is allowed to vary. \texttt{COILOPT++} \cite{Brown2015} improved upon \texttt{COILOPT}, by defining coils using splines, which allows one to straighten modular coils in order to improve access to the plasma. The need for a winding surface was eliminated with the \texttt{FOCUS} \cite{Zhu2018} code, which represents coils as 3-dimensional space curves. The \texttt{FOCUS} approach employs analytic differentiation for gradient-based optimization, as we do in this work. As the design of optimal coils is central to the development of an economical stellarator, it is important to have several approaches. The current potential method could have several possible advantages, including the possible implementation of adjoint methods. Furthermore, the complexity of the nonlinear optimization is reduced over other approaches, as the current distribution on the winding surface is efficiently and robustly computed by solving a linear system. By optimizing the winding surface it is possible to gain insight into what features of plasma surfaces require coils to be close to the plasma, and what features allow coils to be placed farther away \cite{Landreman2016}. 

Many engineering design problems can be formulated in terms of the minimization of an objective function with respect to some free parameters. A powerful tool for such problems is gradient-based optimization, which requires knowledge of the sensitivity of the objective function with respect to design parameters. These gradients can be computed by finite differencing the objective function, but the finite step size introduces errors and the step size must be chosen carefully. Also, if the optimization space is very large, finite differencing can be computationally expensive. Although derivative-free optimization techniques exist, they are less efficient than gradient based algorithms, are limited in the types of constraints that can be implemented, and are typically effective only for small problems \cite{Nocedal2006}. Adjoint methods allow for efficient computation of gradients of the objective function with respect to a large number of design parameters. The cost of computing the derivatives in this way scales independently of the number of design parameters and linearly with the number of objective functions. In addition to gradient-based optimization, these derivatives can also be used for uncertainty quantification in scientific computation \cite{Roy2011} or to construct sensitivity maps for visualization of how an objective function changes with respect to normal displacements of a surface \cite{Othmer2008,Othmer2014}. 

Adjoint methods were developed in the 1970s for sensitivity analysis of drag and flow dynamics \cite{Pironneau1974} and have been widely used for shape optimization in the field of aerodynamics and computational fluid dynamics (CFD) \cite{Kuruvila1995,Jameson1998,Anderson1999,Othmer2008,Othmer2014}. Only recently have these methods been used for tokamak physics in the context of fitting model parameters with experimental edge data on ASDEX-Upgrade \cite{Kim2001} and advanced divertor design with plasma edge simulations \cite{Baelmans2017}. As stellarator design requires many  more geometric parameters than tokamak design, adjoint-based optimization could provide a significant reduction to computational cost to this field. 

The design of magnetic resonance imaging (MRI) coils has also benefited from adjoint methods \cite{Jia2014}. MRI gradient coils which lie on a cylindrical winding surface must provide a specified spatial variation in the magnetic field within a region of interest. This inverse problem is often solved with a linear least-squares system by minimizing the squared departure from the desired field at specified points with respect to the current in differential surface elements \cite{Turner1993}. This method is comparable to the \texttt{NESCOIL} \cite{Merkel1987} approach for stellarator coil design. Gradient coil design was improved by the addition of a regularization term related to the integral-squared current density \cite{Forbes2005} or the integral-squared curvature \cite{Forbes2001}, comparable to the \texttt{REGCOIL} approach. The adjoint method is applied to compute the sensitivity of an objective function with respect to the current potential on the winding surface. Here the Biot-Savart law is written in terms of a matrix equation using the least-squares finite element method, and the adjoint of this matrix is inverted to compute the derivatives \cite{Jia2014}. As the adjoint formalism has proven fruitful in this field, we anticipate that it could have similar applications in the closely-related field of stellarator coil design. 

In the sections that follow, we present a new method for design of the coil-winding surface using adjoint-based optimization. An adjoint solve is performed to obtain gradients of several figures of merit, the integral-squared normal magnetic field on the plasma surface and root-mean-squared current density on the winding surface, with respect to the Fourier components describing the coil surface. A brief overview of the \texttt{REGCOIL} approach is given in \ref{section_REGCOIL}. The optimization method and objective function are described in section \ref{sect_opt}. The adjoint method for computing gradients of the objective function is outlined in section \ref{sect_adjoint}. Optimization results for the W7-X and HSX winding surfaces are presented in section \ref{sect_results}. In section \ref{sect_sensitivity} we demonstrate a method for visualization of shape derivatives on the winding surface. We discuss properties of optimized winding surface configurations in section \ref{sect_configopt}. In section \ref{sect_conclusions} we summarize our results and conclude. 

\section{Overview of the REGCOIL system}
\label{section_REGCOIL}

First, we review the problem of determining coil shapes
once the plasma boundary and coil winding surface have been specified. Given the winding surface geometry, our task is to obtain the surface current density, $\bm{K}$. The divergence-free surface current density can be related to a scalar current potential $\Phi$, the stream function for $\bm{K}$,
\begin{gather}
\bm{K} = \bm{n} \times \nabla \Phi.
\end{gather}
Here $\bm{n}$ is the unit normal on the winding surface. The current potential $\Phi$ can be decomposed into single-valued and secular terms,
\begin{gather}
\Phi(\theta, \zeta) = \Phi_{\text{sv}}(\theta,\zeta) + \frac{ G \zeta}{2 \pi} + \frac{I \theta}{2 \pi}.
\end{gather}
Here $\zeta$ is the usual toroidal angle, and $\theta$ is a poloidal angle. The quantities $G$ and $I$ are the currents linking the surface poloidally and toroidally, respectively. The single-valued term ($\Phi_{\text{sv}}$) is determined by solving the \texttt{REGCOIL} system. It is chosen to minimize the primary objective function,
\begin{gather}
\chi^2 = \chi^2_B + \lambda \chi^2_K.
\label{primary_objective}
\end{gather}
Here $\chi^2_B$ is the surface-integrated-squared normal magnetic field on the desired plasma surface,
\begin{gather}
\chi^2_B = \int_{\text{plasma}} d^2 A \, B_n^2.
\label{chi2_B}
\end{gather}
The normal component of the magnetic field on the plasma surface $B_n$ includes contributions from currents in the plasma, current density $\bm{K}$ on the winding surface, and currents in other external coils. The quantity $\chi^2_K$ is the surface-integrated-squared current density on the winding surface,
\begin{gather}
\chi^2_K = \int_{\text{coil}} d^2A \, K^2. 
\end{gather}
Here $K = \abs{\bm{K}}$. Minimization of $\chi^2_B$ by itself ($\lambda=0$) is fundamentally ill-posed, as very different coil shapes can provide almost identical $B_n$ on the plasma surface (for example, oppositely directed currents cancel in the Biot-Savart integral). The addition of $\chi^2_K$ to the objective function is a form of Tikhonov regularization. As we will show, minimization of $\chi^2_K$ also simplifies coil shapes. The formulation in \texttt{REGCOIL} allows for finer control of regularization while improving engineering properties of the coil set over the \texttt{NESCOIL} formulation, which relies on Fourier series truncation for regularization.

The regularization parameter $\lambda$ can be chosen to obtain a target maximum current density $K_{\text{max}}$, corresponding to a minimum tolerable inter-coil spacing. A 1D nonlinear root finding algorithm is typically used for this process. 

The single-valued part of the current potential $\Phi_{\text{sv}}$ is represented using a finite Fourier series,
\begin{gather}
\Phi_{\text{sv}}(\theta,\zeta) = \sum_{j} \Phi_{j} \sin (m_j \theta - n_j \zeta).
\end{gather}
Only a sine series is needed if stellarator symmetry is imposed on the current density ($K(-\theta,-\zeta) = K(\theta,\zeta)$). As the minimization of $\chi^2$ with respect to $\Phi_{j}$ is a linear least-squares problem, it can be solved via the normal equations to obtain a unique solution. The Fourier amplitudes $\Phi_j$ are determined 
by the minimization of $\chi^2$,
\begin{gather}
\partder{\chi^2}{\Phi_j} = \partder{\chi^2_B}{\Phi_j} + \lambda \partder{\chi^2_K}{\Phi_j} = 0,
\label{regcoil_minimization}
\end{gather}
which takes the form of a linear system,
\begin{gather}
\sum_j A_{k,j} \Phi_j = b_k.
\label{forward}
\end{gather}
We will use the notation $\bm{A} \bm{\Phi} = \bm{b}$. Throughout bold-faced type will denote the vector space of basis functions for $\Phi_{\text{sv}}$ unless otherwise noted. For additional details see \cite{Landreman2017}.

\section{Winding surface optimization}
\label{sect_opt}

We use \texttt{REGCOIL} to compute the distribution of current on a fixed, two-dimensional winding surface. To design coil shapes in 3-dimensional space, we modify the winding surface geometry by minimizing an objective function (\ref{objective_function}). This objective function quantifies key physics and engineering properties and is easy to calculate from the \texttt{REGCOIL} solution. Optimal coil geometries are obtained by nonlinear, constrained optimization.

\subsection{Objective function}
The Cartesian components of the winding surface can be decomposed in Fourier harmonics.
\begin{gather}
x = \sum_{m,n} r_{mn}^c \cos(m \theta + n N_p \zeta) \cos (\zeta), \\
y = \sum_{m,n} r_{mn}^c \cos(m \theta + n N_p \zeta) \sin (\zeta), \\
z = \sum_{m,n} z_{mn}^s \sin(m \theta + n N_p \zeta).
\label{Fourier}
\end{gather}
Here $N_p$ is the number of toroidal periods. Stellarator symmetry of the winding surface is assumed ($R(-\theta,-\zeta) = R(\theta,\zeta)$ and $z(-\theta,-\zeta) = -z(\theta,\zeta)$, where $R^2 = x^2 + y^2$). We take the Fourier components of the winding surface, $\Omega = (r_{mn}^c, z_{mn}^s)$, as our optimization parameters and assume a desired plasma surface to be held fixed. Throughout $\Omega$ displayed with a subscript index will refer to a single Fourier component, while in the absence of a subscript it refers to the set of Fourier components. For a given winding surface geometry, $\Omega$, and desired plasma surface, the current potential $\Phi (\Omega)$ can be determined by solving the \texttt{REGCOIL} system to obtain a solution which both reproduces the desired plasma surface with fidelity and maximizes coil-coil distance, as described in section \ref{section_REGCOIL}.

We define an objective function, $f$, which will be minimized with respect to $\Omega$,
\begin{gather}
f(\Omega, \bm{\Phi}(\Omega))  = \chi^2_B(\Omega, \bm{\Phi}(\Omega)) - \alpha_V V_{\text{coil}}^{1/3}(\Omega) + \alpha_S S_p(\Omega) + \alpha_K  \norm{\bm{K}}_2(\Omega, \bm{\Phi}(\Omega)).
\label{objective_function}
\end{gather}
The coefficients $\alpha_V$, $\alpha_S$, and $\alpha_K$ weigh the relative importance of the terms in $f$. We take $\chi^2_B$ (\ref{chi2_B}) as our proxy for the desired physics properties of the plasma surface. The normal magnetic field depends on $\bm{\Phi}$, the single-valued current potential on the surface, and $\Omega$, the geometric properties of the coil-winding surface. The quantity $V_{\text{coil}}$ is the total volume enclosed by the coil-winding surface,
\begin{gather}
V_{\text{coil}} = \int_{\text{coil}} d^3 V.
\end{gather}
We use $V_{\text{coil}}^{1/3}$ as a proxy for the coil-plasma separation. The quantity $S_p$ is a measure of the spectral width of the Fourier series describing the coil-winding surface \cite{Hirshman1985},
\begin{gather}
S_p = \sum_{m,n} m^{p} \left( (r_{mn}^c)^2 + (z_{mn}^s)^2 \right).
\label{spectral_width}
\end{gather}
Smaller values of $S_p$ correspond to Fourier spectra which decay rapidly with increasing $m$. We take advantage of the non-uniqueness of the representation in (\ref{Fourier}) to obtain surface parameterization which are more efficient. There is no unique definition of $\theta$,
and minimization of $S_p$ removes this redundancy.
We use a typical value of $p=2$. The quantity $\norm{\bm{K}}_2$ is the 2-norm of the current density, defined in terms of an area integral over the surface,
\begin{gather}
\norm{\bm{K}}_2 = \left( \frac{\int_{\text{coil}} d^2 A \, \abs{\bm{K}}^2}{A_{\text{coil}}}  \right)^{1/2},
\end{gather}
where $A_{\text{coil}}$ is the winding surface area,
\begin{gather}
A_{\text{coil}} = \int_{\text{coil}} d^2 A \, .
\end{gather}
Although we are using a current potential approach rather than directly optimizing coil shapes, including $\norm{\bm{K}}_2$ in the objective function allows us to obtain coils with good engineering properties. The direct targeting of coil metrics (such as the curvature) introduces additional arbitrary weights in the objective function, and the solution to another adjoint equation must be obtained to compute its gradient. This will be left for future work. 

\begin{figure}
\includegraphics[width=0.9\textwidth]{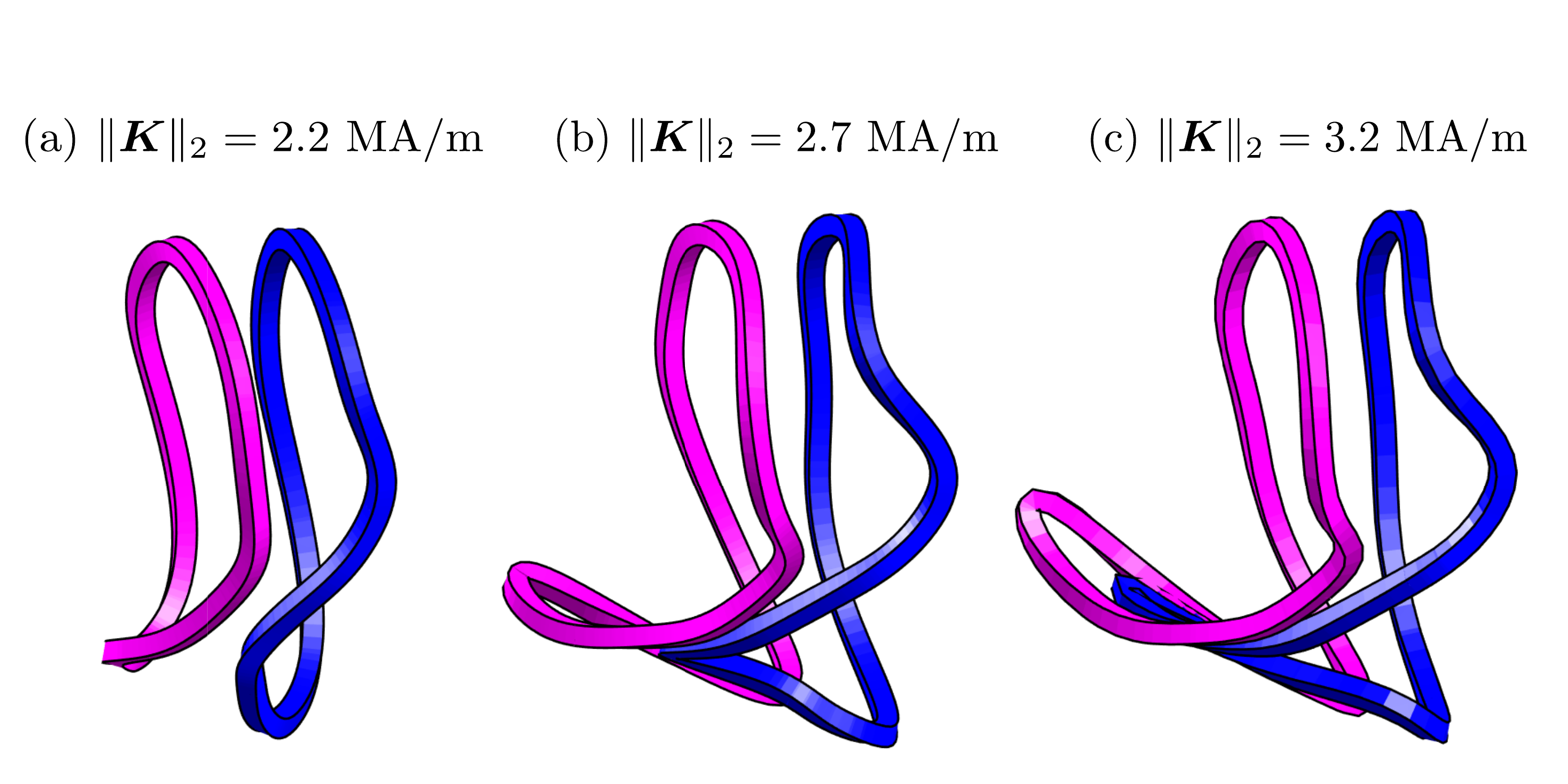}
\caption{Two non-planar W7-X coils (corresponding to the two leftmost coils in figure \ref{w7x_coils}) computed with \texttt{REGCOIL} using the actual W7-X winding surface. The regularization parameter $\lambda$ is chosen to achieve the shown values of $\norm{\bm{K}}_2$. As $\norm{\bm{K}}_2$ increases, the average length, toroidal extent, and curvature increase.}
\label{rmsKcoilcompare}
\end{figure}

To demonstrate this correlation between $\norm{\bm{K}}_2$
and coil shape complexity, we compute the coil set on the actual W7-X winding surface using \texttt{REGCOIL}. The regularization parameter $\lambda$ is varied to achieve several values of $\norm{\bm{K}}_2$. Coil shapes are obtained from the contours of $\Phi$. In figure \ref{rmsKcoilcompare}, two of the W7-X non-planar computed in this way are shown, and the corresponding coil metrics are given in table \ref{rmsKcoilmetrics}. These correspond to the two leftmost coils in figure \ref{w7x_coils}. We consider the average and maximum length $l$, toroidal extent $\Delta \zeta$, and curvature $\kappa$ and the minimum coil-coil distance $d_{\text{coil-coil}}^{\text{min}}$. The average, maximum, and minimum are taken over the set of 5 unique coils.
The coil shapes become more complex as $\norm{\bm{K}}_2$ increases, quantified by increasing $\kappa$ and $\Delta \zeta$ and decreasing $d_{\text{coil-coil}}^{\text{min}}$. 
Here the curvature, $\kappa$, of a 3-dimensional parameterized curve, $\bm{r}(t)$, is 
\begin{gather}
\kappa = \left. \bigg \rvert \der{\bm{r}}{t} \times \der{^2 \bm{r}}{t^2} \bigg \rvert \middle/ \bigg \rvert \der{\bm{r}}{t} \bigg \rvert^3 \right. .
\label{curvature_of_curve}
\end{gather}
We have compared coil shapes on a single winding surface, finding them to become simpler as $\norm{\bm{K}}_2$ decreases. As $\norm{\bm{K}}_2 = \left(\chi^2_K/A_{\text{coil}}\right)^{1/2}$, we would find similar trends with $\chi^2_K$. We have chosen to include $\norm{\bm{K}}_2$ in the objective function as it is normalized by $A_{\text{coil}}$, so it is a more useful quantity for comparison of coil shapes on different winding surfaces. 

\begin{table} 
\renewcommand{\arraystretch}{1.4}
\begin{tabular} { | c | c | c | c |}
\hline
$\norm{\bm{K}}_2$ [MA/m] & 2.20 & 2.70 & 3.20 \\ \hline
$K_{\text{max}}$ [MA/m] & 4.55 & 9.50 & 29.1 \\
Average $l$ [m] & 8.03 & 9.18 & 9.81 \\
Max $l$ [m] & 8.26 & 10.5 & 11.8 \\
Average $\Delta \zeta$ [rad.] & 0.146 & 0.222 & 0.253\\
Max $\Delta \zeta$ [rad.] & 0.161 & 0.282 & 0.372\\
Average $\kappa$ [m$^{-1}$] & 1.04 & 1.29  & 1.32 \\
Max $\kappa$ [m$^{-1}$] & 2.54 & 20.3 & 56.1 \\
$d_{\text{coil-coil}}^{\text{min}}$ [m] & 0.353 & 0.182 & 0.0758 \\ \hline
\end{tabular}
\caption{Comparison of metrics for coils computed with \texttt{REGCOIL} using the actual W7-X winding surface. Average and max are evaluated for the set of 5 unique coils. The regularization parameter $\lambda$ is varied to achieve these values of $\norm{\bm{K}}_2$.}
\label{rmsKcoilmetrics}
\end{table}

To minimize $f$, the relative weights in (\ref{objective_function}) ($\alpha_V$, $\alpha_S$, and $\alpha_K$) are chosen such that each of the terms in the objective function have similar magnitudes, though much tuning of these parameters is required to obtain results which simultaneously improve the physics properties (decrease $\chi^2_B$) and engineering properties (increase $V_{\text{coil}}$ and $d_{\text{coil-coil}}^{\text{min}}$, decrease $\kappa$ and $\Delta \zeta$).

\subsection{Optimization constraints}
\label{sect_constraint}

Minimization of $f$ is performed subject to the inequality constraint $d_{\text{min}} \geq d_{\text{min}}^{\text{target}}$. Here $d_{\text{min}}$ is the minimum distance between the coil-winding surface and the plasma surface,
\begin{gather}
d_{\text{min}} = \min_{\theta,\zeta} \left( d_{\text{coil-plasma}} \right) = \min_{\theta,\zeta} \left( \min_{\theta_p, \zeta_p} \, \abs{ \bm{r}_{\text{coil}} - \bm{r}_{\text{plasma}} }  \right),
\end{gather}
and $d_{\text{min}}^{\text{target}}$ is the minimum tolerable coil-plasma separation. The quantities $\theta_p$ and $\zeta_p$ are poloidal and toroidal angles on the plasma surface, $\bm{r}_{\text{plasma}}$ and $\bm{r}_{\text{coil}}$ are the position vectors on the plasma and winding surface, and $d_{\text{coil-plasma}}$ is the coil-plasma distance as a function of $\theta$ and $\zeta$. 

The maximum current density $K_{\text{max}}$ is also constrained,
\begin{gather}
K_{\text{max}} = \max_{\theta,\zeta} \, K .
\end{gather}
This roughly corresponds to a fixed minimum coil-coil spacing. This constraint is enforced by fixing $K_{\text{max}}$ to obtain the regularization parameter $\lambda$ in the \texttt{REGCOIL} solve, so we avoid the need for an equality constraint or the inclusion of $K_{\text{max}}$ in the objective function. Rather, $\Phi(\Omega)$ is determined such that $K_{\text{max}}$ is fixed. The inequality-constrained nonlinear optimization is performed using the NLOPT \cite{NLOPT} software package using a conservative convex separable quadratic approximation (CCSAQ) \cite{Svanberg2002}. While there are several gradient-based inequality-constrained algorithms available, we chose to use CCSAQ as it is relatively insensitive to the bound constraints imposed on the optimization parameters. We recognize that there are many possible combinations of constraints, objective functions, and regularization conditions that could be used. For example, $\norm{\bm{K}}_2$ could be fixed to determine $\lambda$ while $K_{\text{max}}$ could be included in the objective function. We found that the formulation we have presented produces the best coil shapes. 

\section{Derivatives of $f$ and the adjoint method}
\label{sect_adjoint}
We must compute derivatives of $f$ with respect to the geometric parameters $\Omega$ in order to use gradient-based optimization methods. The spectral width $S_p$ and volume $V_{\text{coil}}$ are explicit functions of $\Omega$, so their analytic derivatives can be obtained. On the other hand, $\chi^2_B$ and $\norm{\bm{K}}_2$ depend both explicitly on coil geometry and on $\bm{\Phi}(\Omega)$. One approach to obtain the derivatives of these quantities could be to solve the \texttt{REGCOIL} linear system $N_{\Omega} +1$ times, taking a finite difference step in each Fourier coefficient. However, if $N_{\Omega}$ (number of Fourier modes) is large, the computational cost of this method could be prohibitively expensive. Instead we will apply the adjoint method to compute derivatives. This technique will be demonstrated below. 

The derivative of $\chi^2_B$ can be computed using the chain rule, 
\begin{gather}
\partder{\chi^2_B(\Omega, \bm{\Phi}(\Omega))}{\Omega_j} \bigg \rvert_{\bm{A} \bm{\Phi} = \bm{b}} = \partder{\chi^2_B}{\Omega_j} \bigg \rvert_{\bm{\Phi}} + \partder{\chi^2_B}{\bm{\Phi}} \cdot \partder{\bm{\Phi}}{\Omega_j} \bigg \rvert_{\bm{A} \bm{\Phi} = \bm{b}}.
\label{sensitivity_der}
\end{gather}
The subscript $\bm{A} \bm{\Phi} = \bm{b}$ indicates that $\bm{\Phi}$ varies with $\Omega$ according to (\ref{forward}), with $\bm{A}$ and $\bm{b}$ denoting the matrix and right hand side of the linear system in (\ref{forward}). The dot product is a contraction over the current potential basis functions, $\Phi_j$. We can compute $\partial \bm{\Phi}/ \partial \Omega_j$ by differentiating the linear system (\ref{forward}) with respect to $\Omega_j$, 
\begin{gather}
\partder{\bm{A}}{\Omega_j} \bm{\Phi} + \bm{A} \partder{\bm{\Phi}}{\Omega_j} = \partder{\bm{b}}{\Omega_j},
\end{gather}
and formally solving this equation to obtain
\begin{gather}
\partder{\bm{\Phi}}{\Omega_j} = \bm{A}^{-1} \left( \partder{\bm{b}}{\Omega_j} - \partder{\bm{A}}{\Omega_j} \bm{\Phi} \right).
\label{linear_sensitivity}
\end{gather}
Equation (\ref{linear_sensitivity}) is inserted into  (\ref{sensitivity_der}),
\begin{gather}
\partder{\chi^2_B(\Omega, \bm{\Phi}(\Omega))}{\Omega_j} \bigg \rvert_{\bm{A} \bm{\Phi} = \bm{b}} = \partder{\chi^2_B}{\Omega_j} \bigg \rvert_{\bm{\Phi}} + \partder{\chi^2_B}{\bm{\Phi}} \cdot \left[ \bm{A}^{-1} \left( \partder{\bm{b}}{\Omega_j} - \partder{\bm{A}}{\Omega_j} \bm{\Phi} \right) \right].
\end{gather}
This expression could be evaluated by inverting $\bm{A}$ for each of the geometric components $\Omega_j$ and performing the inner product with $\partial \chi^2_B/ \partial \bm{\Phi}$ for each $\Omega_j$. However, the computational cost of this method scales similarly to that of finite differencing. 
Instead, we can exploit the adjoint property of the operator. For a given inner product $(\,\,,\,\,)$, the adjoint of an operator, $A$, is defined as the operator $A^{\dagger}$ satisfying $(b,Ac) = (A^{\dagger} b, c)$. As we are working in $\mathbb{R}^n$, the adjoint operator corresponds to the matrix transpose, so
\begin{gather}
\partder{\chi^2_B(\Omega, \bm{\Phi}(\Omega))}{\Omega_j} \bigg \rvert_{\bm{A} \bm{\Phi} = \bm{b}} = \partder{\chi^2_B}{\Omega_j} \bigg \rvert_{\bm{\Phi}} + \left[ \left(\bm{A}^{-1}\right)^{T} \partder{\chi^2_B}{\bm{\Phi}}\right] \cdot \left( \partder{\bm{b}}{\Omega_j} - \partder{\bm{A}}{\Omega_j} \bm{\Phi} \right).
\end{gather}
For any invertible matrix, $\left( \bm{A}^{-1} \right)^T = \left( \bm{A}^{T} \right)^{-1}$. Hence we can instead invert the operator $\bm{A}^T$ to compute an adjoint variable $\bm{q}$, defined as the solution of
\begin{gather}
\bm{A}^T \bm{q} = \partder{\chi^2_B}{\bm{\Phi}}.
\label{adjoint}
\end{gather}
Rather than finite difference in each $\Omega_j$ or invert $\bm{A}$ for each $\partial \bm{\Phi}/\partial \Omega_j$ as in (\ref{linear_sensitivity}), we solve two linear systems: the forward (\ref{forward}) and adjoint (\ref{adjoint}). The adjoint equation is similar to the forward equation ($\bm{A}^T$ has the same dimensions and eigenspectrum as $\bm{A}$), so the same computational tools can be used to solve the adjoint problem. We then perform an inner product with $\bm{q}$ to obtain the derivatives with respect to each $\Omega_j$,
\begin{gather}
\partder{\chi^2_B(\Omega, \bm{\Phi}(\Omega))}{\Omega_j} \bigg \rvert_{\bm{A} \bm{\Phi} = \bm{b}} = \partder{\chi^2_B}{\Omega_j} \bigg \rvert_{\bm{\Phi}} + \bm{q} \cdot \left( \partder{\bm{b}}{\Omega_j} - \partder{\bm{A}}{\Omega_j} \bm{\Phi} \right).
\label{adjointsensitivity}
\end{gather}
The derivatives $\partial \bm{b}/\partial \Omega_j$, $\partial \bm{A}/\partial \Omega_j$, $\left( \partial \chi^2_B/\partial \Omega_j \right)_{\bm{\Phi}}$, and $\partial \chi^2_B/\partial \bm{\Phi}$ can be computed analytically. In the above discussion, the regularization parameter $\lambda$ has been assumed to be fixed. A similar method can be used if a $\lambda$ search is performed to obtain a target $K_{\text{max}}$ (see appendix \ref{lambda_search}). The same method is used to compute derivatives of $\norm{\bm{K}}_2$. 

We note that adjoint methods provide the most significant reduction in computational cost when the linear solve is expensive. For the \texttt{REGCOIL} system this is not the case, as the cost of constructing $\bm{A}$ and $\bm{b}$ exceeds that of the solve. We have implemented OpenMP multithreading for the construction of $\partial \bm{A}/\partial \Omega$ and $\partial \bm{b}/\partial \Omega$ such that the cost of computing the gradients via the adjoint method is cheaper than computing finite differences serially.

The constraint functions, $d_{\text{min}}$ and $K_{\text{max}}$, must also be differentiated with respect to $\Omega_j$. As $d_{\text{min}}$ is defined in terms of the minimum function, we approximate it using the smooth log-sum-exponent function \cite{Boyd2004}.
\begin{gather}
d_{\text{min, lse}} = - \frac{1}{q} \log \left( \frac{\int_{\text{coil}} d^2 A \, \int_{\text{plasma}} d^2 A \, \exp \left( - q \abs{\bm{r}_{\text{coil}} - \bm{r}_{\text{plasma}}} \right) }{\int_{\text{coil}} d^2 A \, \int_{\text{plasma}} d^2 A \, } \right)
\label{lse_d}
\end{gather}
This function can be analytically differentiated with respect to $\Omega_j$. As $q$ approaches infinity, $d_{\text{min, lse}}$ approaches $d_{\text{min}}$. For $q$ very large, the function obtains very sharp gradients. A typical value of $q = 10^4$ m$^{-1}$ was used. The log-sum-exponent function is also used to approximate $K_{\text{max}}$. 

\FloatBarrier
\section{Winding surface optimization results}
\label{sect_results}
\FloatBarrier

\subsection{Trends with optimization parameters}

\begin{figure}
\includegraphics[width=.8\textwidth]{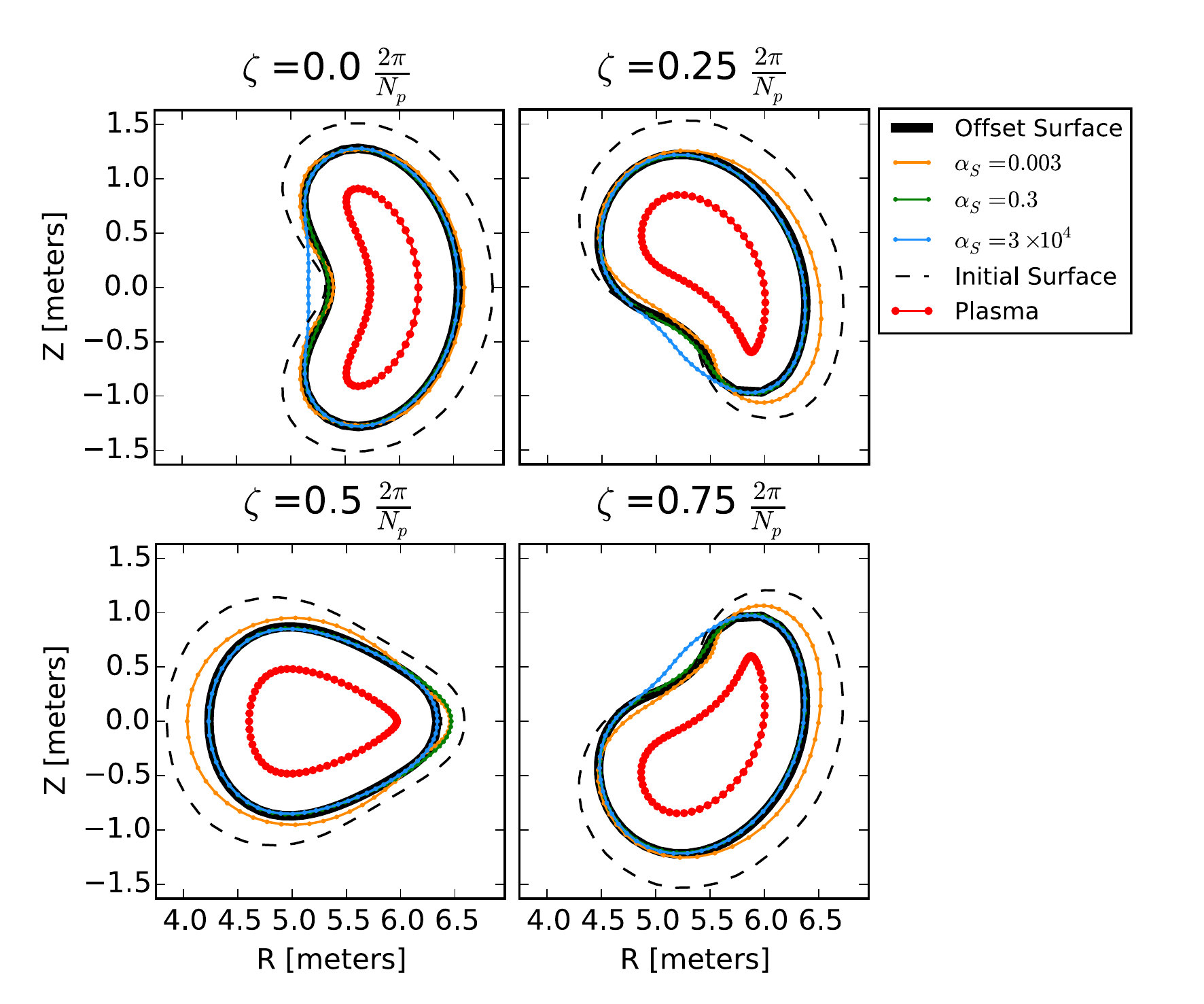}
\caption{Optimized winding surfaces obtained with $\alpha_V = \alpha_K = 0$ and the values of $\alpha_S$ shown. The actual W7-X winding surface is used as the initial surface in the optimization (black dashed). As $\alpha_S$ increases, the magnitude of the spectral-width term in the objective function increases, and the winding surface approaches a cylindrical torus with a minimal Fourier spectra. For moderately small values of $\alpha_S$, the winding surface approaches a uniform offset surface from the plasma surface (black solid).}
\label{alpha2_scan}
\end{figure}

\begin{figure}
\includegraphics[width=.8\textwidth]{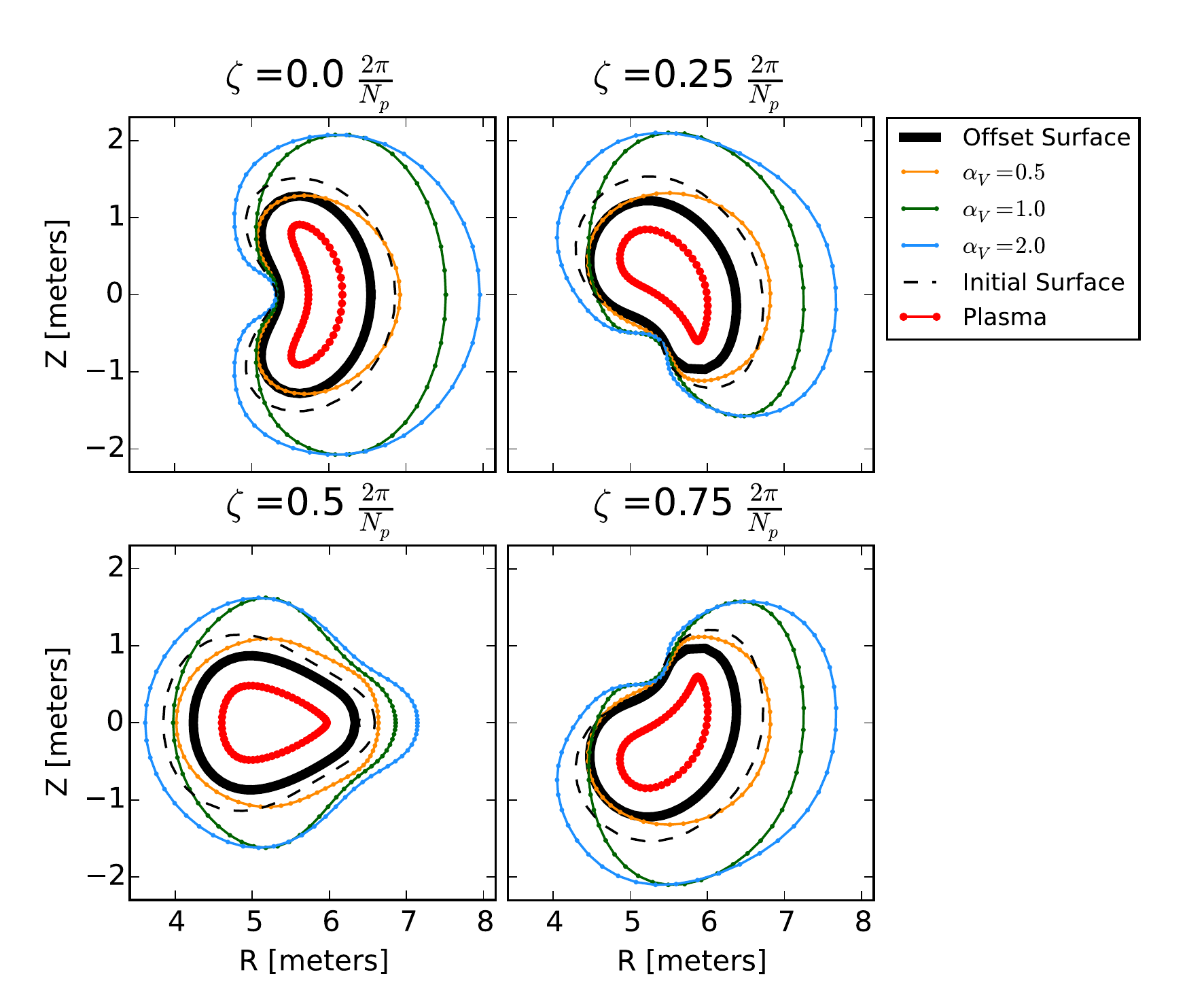}
\caption{Optimized winding surfaces obtained with $\alpha_S = 0.3$, $\alpha_K = 0$, and the values of $\alpha_V$ shown. The actual W7-X winding surface is used as the initial surface in the optimization (black dashed). As $\alpha_V$ increases, $d_{\text{coil-plasma}}$ increases on the outboard side while it remains fixed in the concave region.}
\label{alpha1_scan}
\end{figure}

Beginning with the actual W7-X winding surface, we perform scans over the coefficients $\alpha_V$ and $\alpha_S$ in the objective function (\ref{objective_function}). The plasma surface was obtained from a fixed-boundary VMEC solution that predated the coil design and is free from modular coil ripple. The constraint target is set to be the minimum coil-plasma distance on the initial winding surface, $d_{\text{min}}^{\text{target}}=0.37$ m. The cross sections of the optimized surfaces in the poloidal plane are shown in figures \ref{alpha2_scan} and \ref{alpha1_scan} along with the last-closed flux surface (red), a constant offset surface at $d_{\text{min}}^{\text{target}}$ (black solid), and the initial winding surface (black dashed).

With increasing $\alpha_S$ at fixed $\alpha_V = \alpha_K = 0$, the winding surface approaches a cylindrical torus which has a minimal Fourier spectra. At moderately small values of $\alpha_S$ (0.3) the surface approaches a constant offset surface at $d_{\text{min}}^{\text{target}}$, as $\chi^2_B$ is dominant in objective function. For very small values of $\alpha_S$ (0.003), we find that the optimization terminates at a point relatively close to the initial surface, and the resulting winding surface deviates from a constant offset surface. An intermediate value of $\alpha_S = 0.3$ was chosen for the following optimizations of the W7-X winding surface. 

A scan over $\alpha_V$ is performed at fixed $\alpha_S = 0.3$ and $\alpha_K = 0$ such that the spectral width does not greatly increase. As $\alpha_V$ increases, $d_{\text{coil-plasma}}$ increases significantly on the outboard side while it remains fixed in the inboard concave regions. This trend is not surprising, as concave plasma shapes have been shown to be inefficient to produce with coils \cite{Landreman2016}. 
Interestingly, the winding surface obtains a somewhat pointed shape at the triangle cross-section ($\zeta = 0.5$ $2\pi/N_p$), becoming elongated at the tip of the triangle and `pinching' toward the plasma surface at the edges. 

\subsection{Optimal W7-X winding surface}
\label{w7x_results}

\begin{figure}
\includegraphics[width=1\textwidth]{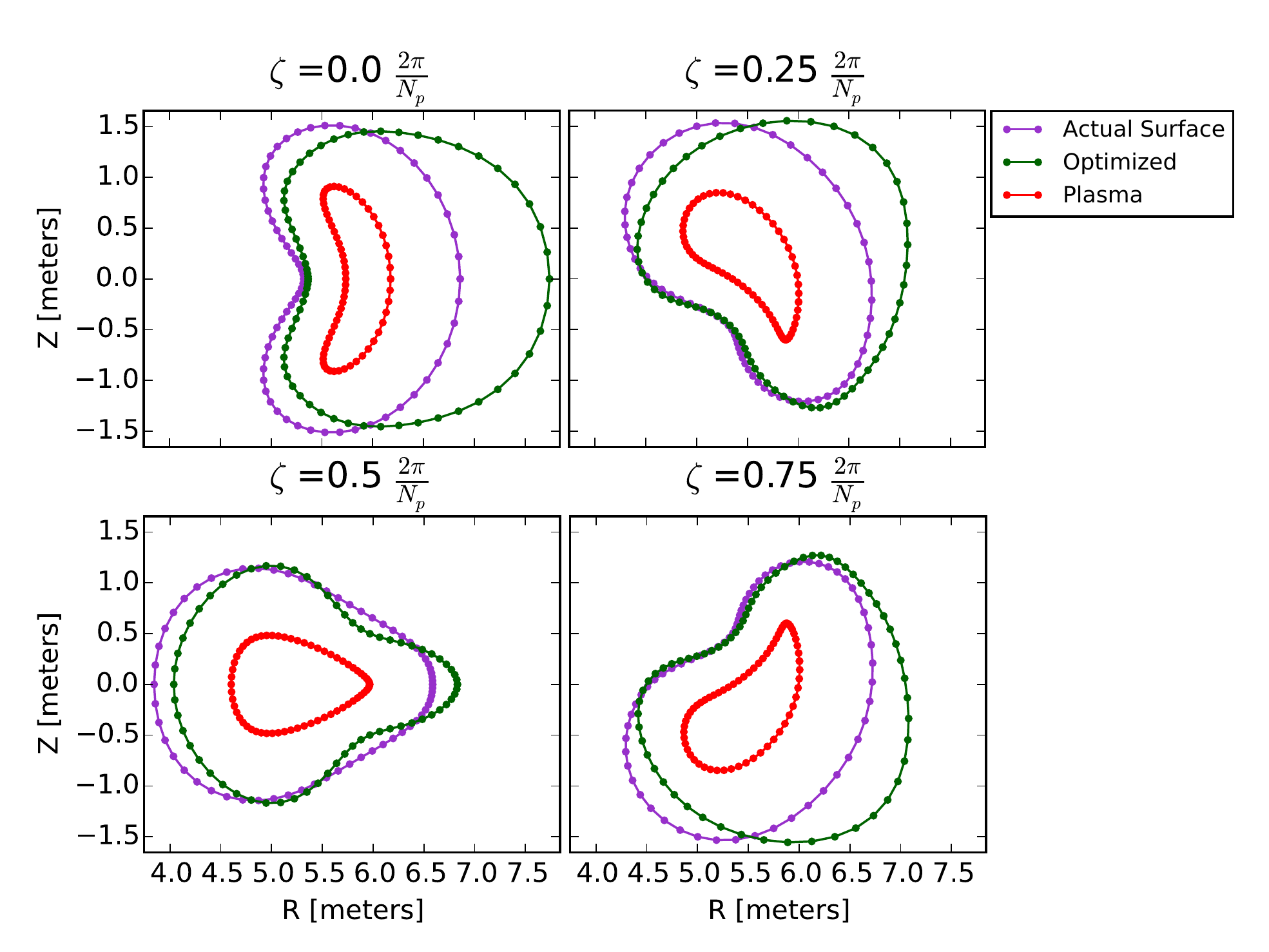}
\caption{The actual W7-X coil-winding surface and plasma surface are shown with our optimized winding surface. In comparison with the actual surface, the optimized surface reduced $\chi^2_B$ by 52\% and increased $V_{\text{coil}}$ by 22\%.}
\label{w7x_surf}
\end{figure}

\begin{figure}
\includegraphics[width=0.8\textwidth]{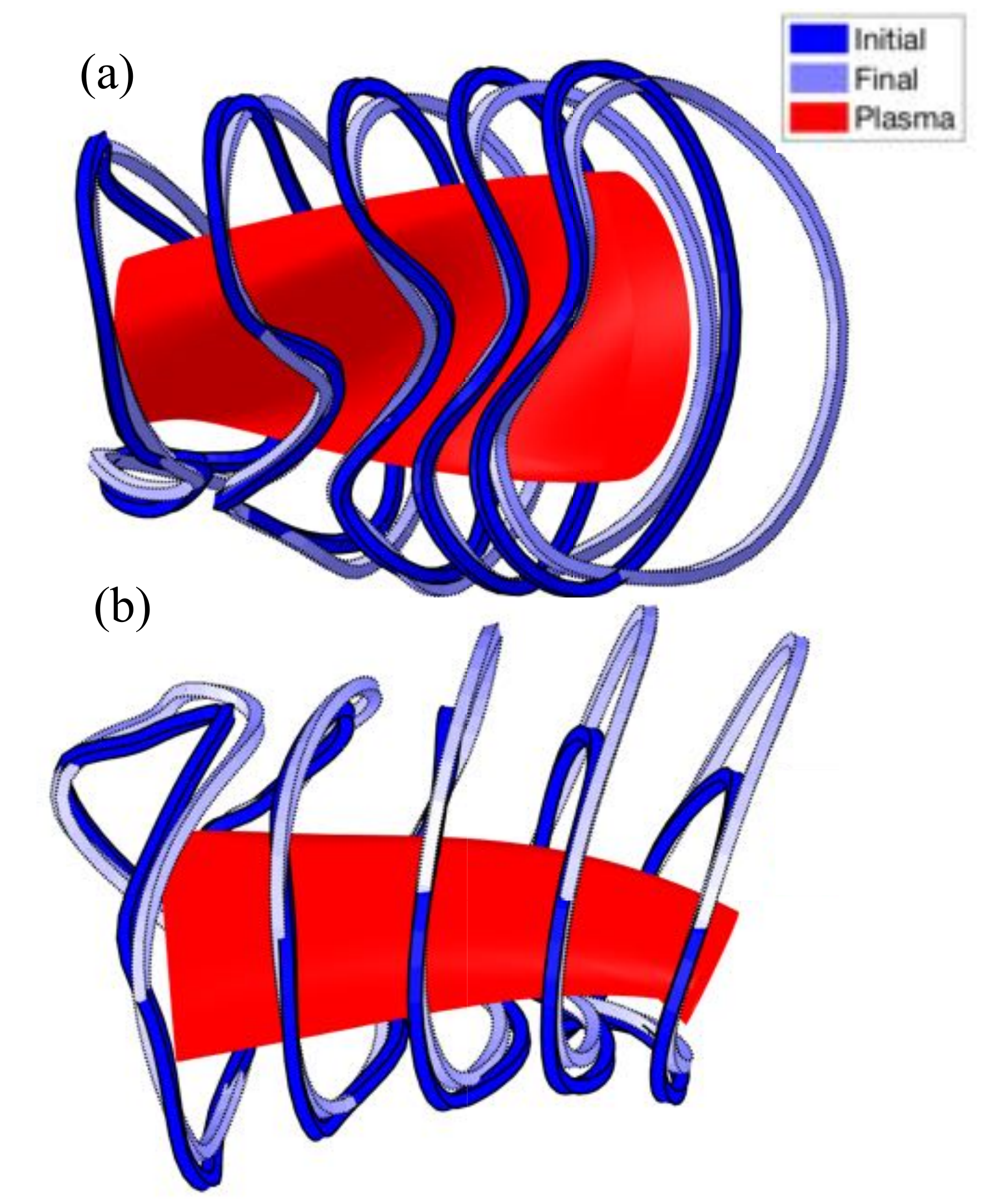}
\caption{Comparisons of coil set computed with \texttt{REGCOIL} using the actual W7-X winding surface (dark blue) and the optimized surface (light blue).}
\label{w7x_coils}
\end{figure}

We now include nonzero $\alpha_K$ and attempt a comprehensive optimization. The $K_{\text{max}}$ constraint is selected such that the metrics ($l$, $\kappa$, and $\Delta \zeta$) of the coils computed on the initial surface roughly match those of the actual non-planar coil set. The coil-plasma distance constraint  $d_{\text{min}}^{\text{target}}$ is set to be the minimum $d_{\text{coil-plasma}}$ on the initial winding surface. Parameters $\alpha_V = 0.5$, $\alpha_S = 0.24$, and $\alpha_K = 1.6\times10^{-6}$ were used in the objective function. Optimization was performed over 118 Fourier coefficients $\big(\abs{n} \leq 4$ and $m \leq 6$ in (\ref{Fourier})$\big)$ and the objective function was evaluated a total of 5165 times to reach the optimum ($1.5\times 10^4$ linear solves rather than $6.1\times10^5$ required for finite difference derivatives). The optimal surface and coil set are shown in figures \ref{w7x_surf} and \ref{w7x_coils}, and the corresponding metrics are shown in table \ref{table_w7x}. We find a solution which increases $V_{\text{coil}}$ by 22\% and decreases $\chi^2_B$ by 52\% over the initial winding surface (note that it is numerically impossible to obtain a current distribution that exactly reproduces the plasma surface, so $\chi^2_B$ is nonzero when computed from the \texttt{REGCOIL} solution on the initial winding surface). In addition, the optimized coil set features a smaller average and maximum $\Delta \zeta$ and $\kappa$ and larger $d_{\text{coil-coil}}^{\text{min}}$. The length of the coils increases to accommodate for the increase in $V_{\text{coil}}$. Again we find that the increase in $V_{\text{coil}}$ is most pronounced in the outboard convex regions while $d_{\text{coil-plasma}}$ is maintained in the concave regions of the bean-shaped cross-sections. The `pinching' feature of the winding surface is again present in the triangle cross-section ($\zeta = 0.5 \, 2\pi/N_p$).  

It should be noted that the decrease in $d_{\text{coil-plasma}}$ at the bottom and top of the bean cross section ($\zeta = 0$) might interfere with the current W7-X divertor baffles. However, the increase in volume on the outboard side would allow for increased flexibility for the neutral beam injection duct \cite{Rust2011}. We have performed this optimization to show that a winding surface could be constructed which increases $V_{\text{coil}}$ (and thus the average $d_{\text{coil-plasma}}$), improves coil shapes, and decreases $\chi^2_B$. If further engineering considerations were necessary these could be implemented. 

\begin{table} 
\renewcommand{\arraystretch}{1.4}
\begin{tabular} {| c | c | c | c |}
\hline
 & Initial & Optimized & Actual coil set \\ \hline
$\chi^2_B$ [T$^2$m$^2$] & 0.115 & 0.0711 & \\
$V_{\text{coil}}$[m$^3$] & 156 & 190 & \\
$\norm{\bm{K}}_2$ [MA/m] & 2.21 & 2.16 & \\
$K_{\text{max}}$ [MA/m] & 7.70 & 7.70 & \\
Average $l$ [m] & 8.51 & 8.95 & 8.69  \\
Max $l$ [m] & 8.84 & 9.14 & 8.74 \\
Average $\Delta \zeta$ [rad.] & 0.190 & 0.179 & 0.198 \\
Max $\Delta \zeta$ [rad.] & 0.222 & 0.197 & 0.208 \\
Average $\kappa$ [m$^{-1}$] & 1.21 & 1.10 & 1.20 \\
Max $\kappa$ [m$^{-1}$] & 9.01 & 4.84 & 2.59 \\
$d_{\text{coil-coil}}^{\text{min}}$ [m] & 0.223 & 0.271 & 0.261 \\ \hline
\end{tabular}
\caption{Comparison of metrics of the actual W7-X winding surface and our optimized surface. We also show metrics of the coil set computed on the winding surfaces using \texttt{REGCOIL} and the metrics for the actual W7-X nonplanar coils. Regularization in \texttt{REGCOIL} is chosen such that the coil metrics computed on the initial surface roughly match those of the actual coil set. Coil complexity improves from the initial to the final surface (decreased average and max $\Delta \zeta$ and $\kappa$, increased $d_{\text{coil-coil}}^{\text{min}}$). The average and max $l$ increases to allow for the increase in $V_{\text{coil}}$.}
\label{table_w7x}
\end{table}

\subsection{Optimal HSX winding surface}
\label{hsx_results}

\begin{figure}
\includegraphics[width=1\textwidth]{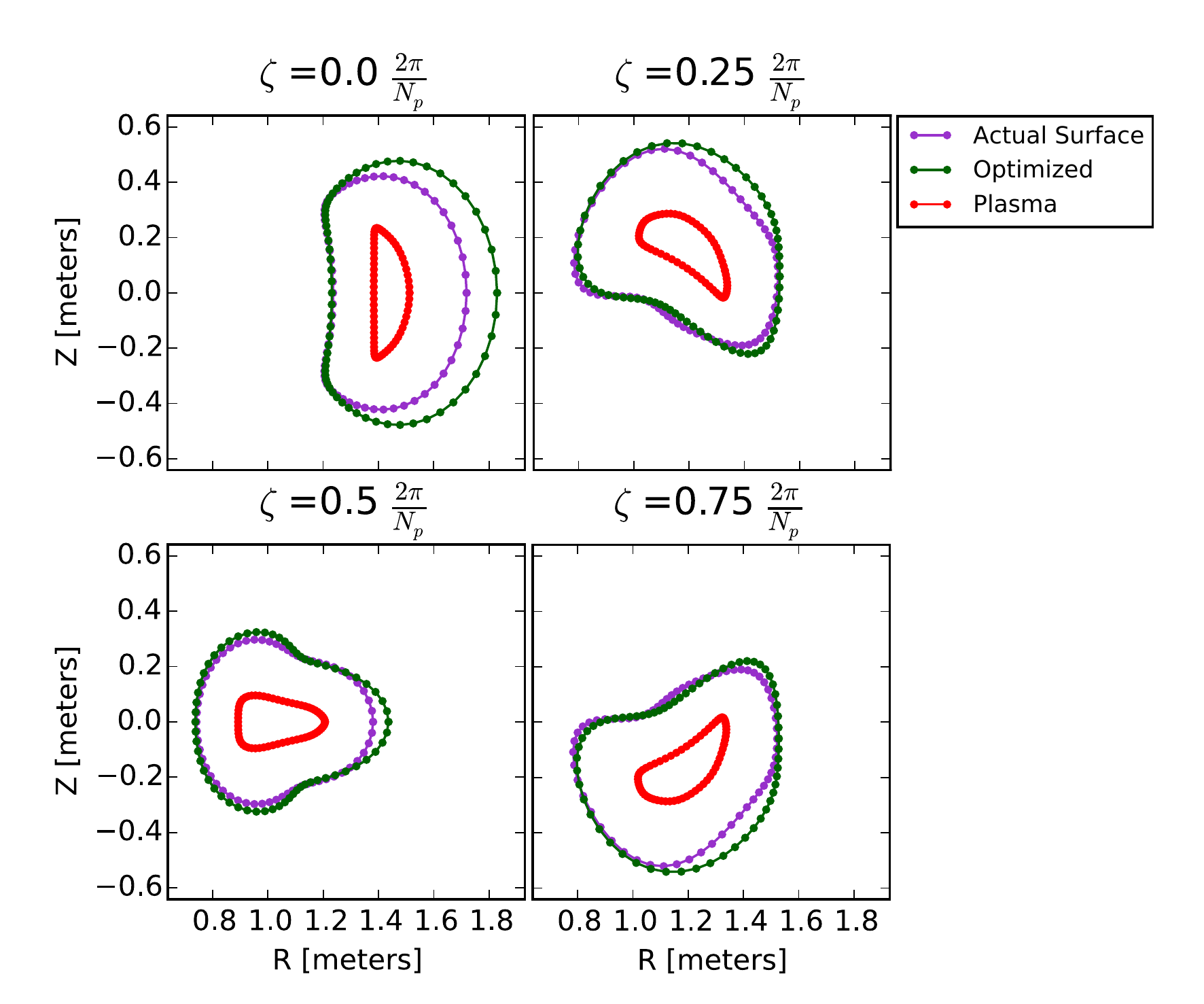}
\caption{The actual HSX coil-winding surface and plasma surface are shown with our optimized winding surface. In comparison with the actual surface, the optimized surface has decreased $\chi^2_B$ by 4\% and increased $V_{\text{coil}}$ by 18\%.}
\label{hsx_surf}
\end{figure}

\begin{figure}
\includegraphics[width=0.8\textwidth]{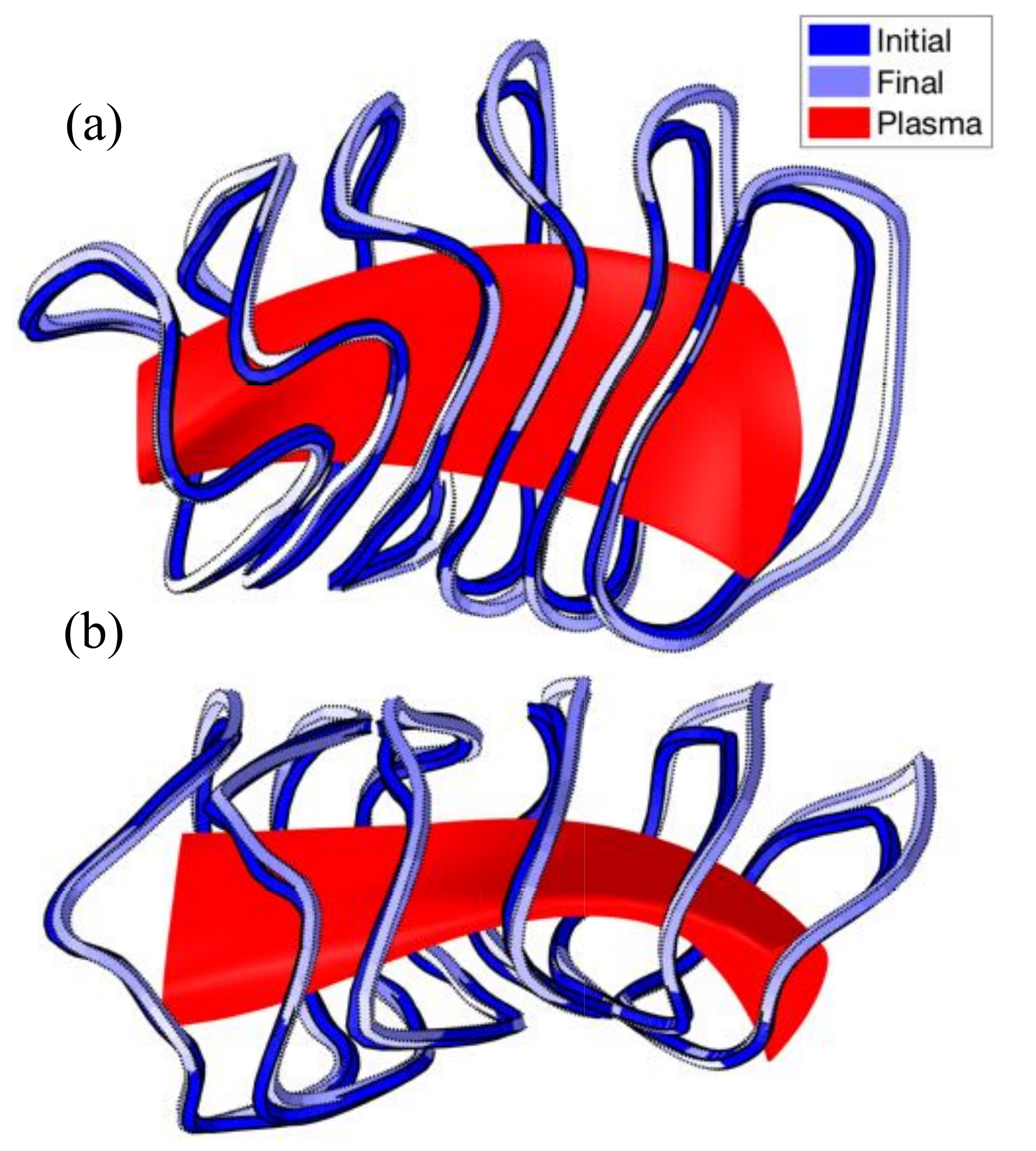}
\caption{The coils obtained from \texttt{REGCOIL} using the actual HSX winding surface (dark blue) and optimized surface (light blue).}
\label{hsx_coils}
\end{figure}

We perform the same procedure for optimization of the HSX winding surface. Parameters $\alpha_V = 3.13\times 10^{-4}$, $\alpha_S = 0$, and $\alpha_K = 3\times 10^{-10}$ were used in the objective function. We found that the spectral width term was not necessary to obtain a satisfying optimum in this case. The initial winding surface was taken to be a toroidal surface on which the actual modular coils lie. The plasma equilibrium used is a fixed-boundary VMEC solution without coil ripple. Optimization was performed over 100 Fourier coefficients $\big(\abs{n} \leq 5$ and $m \leq 4$ in (\ref{Fourier})$\big)$ and the objective function was evaluated a total of 560 times to reach the optimum ($1.7\times10^3$ linear solves rather than $5.7\times10^4$ required for finite difference derivatives). The coil-plasma distance constraint was set to be $d_{\text{min}}^{\text{target}} = 0.14$ m, the minimum coil-plasma distance on the actual winding surface. The optimal surface and coil set are shown in figures \ref{hsx_surf} and \ref{hsx_coils}, and the corresponding coil metrics are shown in table \ref{table_hsx}. We find a solution which increases $V_{\text{coil}}$ by 18\% and decreases $\chi^2_B$ by 4\% over the initial winding surface. The coil set computed with \texttt{REGCOIL} using the optimized surface appears qualitatively similar to that computed with the initial surface but with increased $d_{\text{coil-plasma}}$ on the outboard side. The average and maximum $\Delta \zeta$ and $\kappa$ decreased while $d_{\text{coil-coil}}^{\text{min}}$ was increased for the coil set computed on the optimal surface in comparison to that of the initial surface. As was observed in the W7-X optimization (figure \ref{w7x_surf}), the optimized HSX winding surface obtains a somewhat pinched shape near the triangle cross-section ($\zeta = 0.5 \, 2\pi/N_p$). 

\begin{table} 
\renewcommand{\arraystretch}{1.4}
\begin{tabular} {| c | c | c | c |}
\hline
& Initial & Optimized & Actual coil set \\ \hline
$\chi^2_B$ [T$^2$m$^2$] & $1.53 \times 10^{-5}$ & $1.47\times 10^{-5}$ & \\
$V_{\text{coil}}$[m$^3$] & 2.60 & 3.07 & \\
$\norm{\bm{K}}_2$ [MA/m] & 0.956 & 0.891 & \\
$K_{\text{max}}$ [MA/m] & 1.84 & 1.84 & \\
Average $l$ [m] & 2.26 & 2.39 & 2.24  \\
Max $l$ [m] & 2.49 & 2.46 & 2.33 \\
Average $\Delta \zeta$ [rad.] & 0.372 & 0.365 & 0.362 \\
Max $\Delta \zeta$ [rad.] & 0.530 & 0.505 & 0.478 \\
Average $\kappa$ [m$^{-1}$] & 5.15 & 4.80 & 5.05 \\
Max $\kappa$ [m$^{-1}$] & 33.4 & 25.8 & 11.7 \\
$d_{\text{coil-coil}}^{\text{min}}$ [m] & 0.0850 & 0.0853 & 0.0930 \\ \hline
\end{tabular}
\caption{Comparison of metrics of the actual HSX winding surface and our optimized surface. We also show metrics of the coil set computed on the winding surfaces using \texttt{REGCOIL} and the metrics for the actual HSX modular coils. Regularization in \texttt{REGCOIL} is chosen such that the coil metrics computed on the initial surface roughly match those of the actual coil set. Coil complexity improves from the initial to the final surface (decreased  average and max $\Delta \zeta$ and $\kappa$, increased $d_{\text{coil-coil}}^{\text{min}}$). The average and max $l$ increases to allow for the increase in $V_{\text{coil}}$.}
\label{table_hsx}
\end{table}

\FloatBarrier
\section{Winding surface sensitivity maps}
\label{sect_sensitivity}
\FloatBarrier

With the adjoint method we have computed derivatives of the objective function with respect to Fourier components of the winding surface, $\partial f/\partial \Omega$. While this representation of derivatives is convenient for gradient-based optimization, visualization of the surface sensitivity in real space is obscured. Alternatively, it is possible to represent the sensitivity of $f$ with respect to normal displacements of surface area elements of a given winding surface $\Omega$,
\begin{gather}
\delta f(\Omega, \delta \bm{r}) = \int_{\text{coil}} d^2 A \, S \, \delta \bm{r} \cdot \bm{n}. 
\label{shapederivative}
\end{gather}
Here, $S(\theta,\zeta) $ is a scalar function that will be called the sensitivity. The form (\ref{shapederivative}) implies $f$ is unchanged by tangential displacements of the surface. The shape derivative $\delta f$ can be formally defined as follows \cite{Delfour2011}. Consider a vector field, $\delta \bm{r}$, which describes displacements of the surface, $\Omega$. The surface varies smoothly from $\Omega$ to $\Omega_{\epsilon}$, where each point on $\Omega$ undergoes transformation $T_{\epsilon}$. 
\begin{gather}
\Omega_{\epsilon} = \left \{ T_{\epsilon}(\bm{r}_0) : \bm{r}_0 \in \Omega \right \},
\end{gather}
and $T_{\epsilon}$ is the displacement of each point on the surface by the vector field $\epsilon \delta \bm{r}$,
\begin{gather}
T_{\epsilon}(\bm{r}) = \bm{r} + \epsilon \delta \bm{r}(\bm{r}).
\end{gather}
The shape derivative, $\delta f(\Omega, \delta \bm{r})$, of a functional of the surface geometry, $f(\Omega)$, is then defined as
\begin{gather}
\delta f(\Omega, \delta \bm{r}) = \lim_{\epsilon \rightarrow 0} \frac{ f(\Omega_{\epsilon}) - f(\Omega)}{\epsilon}.
\end{gather}
Note that the definition of $\delta f$ only depends on the direction of $\delta \bm{r}$, not its magnitude. The shape derivative is a G\^{a}teaux derivative, a directional derivative defined for a functional of a vector space. At each point on the winding surface $\delta f(\Omega)$ is defined for each direction $\delta{\bm{r}}$, corresponding to perturbations of the surface at that location in the specified direction. 
Under some assumptions, the shape derivative can be represented in the form of (\ref{shapederivative}) 
(called the Hadamard-Zol\`{e}sio structure theorem by
some authors)
\cite{Delfour2011}. This so-called Hadamard form for shape derivatives is convenient for computation and has been applied to construct sensitivity maps of Navier-Stokes flows for car aerodynamic design \cite{Othmer2008,Othmer2014}. This representation could have potential applications for stellarator design, allowing for visualization of regions on the winding surface which require tight engineering tolerances for a given figure of merit. 

As both $\chi^2_B$ and $\norm{\bm{K}}_2$ are defined in terms of surface integrals over the winding surface, it can be shown that the shape derivative of these functions 
can be written the Hadamard form \cite{Novotny2013}. The surface sensitivity functions $S_{\chi^2_B}$ and $S_{\norm{\bm{K}}_2}$ can be computed from the Fourier derivatives ($\partial \chi^2_B/\partial \Omega$ and $\partial \norm{\bm{K}}_2/\partial \Omega$) using a singular value decomposition method \cite{Landreman2018}. Here the perturbations $\delta f$ and $\delta \bm{r}$ are written in terms of the Fourier derivatives, and $S$ is also represented in a finite Fourier series,
\begin{gather}
\partder{f}{\Omega_j} = \int_{\text{coil}} d^2 A \, \left( \sum_{mn} S_{mn} \cos(m\theta+n N_p \zeta) \right) \partder{\bm{r}}{\Omega_j} \cdot \bm{n}.
\label{SVD_sensitivity}
\end{gather}
After discretizing in $\theta$ and $\zeta$, (\ref{SVD_sensitivity}) takes the form of a (generally not square) matrix equation which can be solved using the Moore-Penrose pseudoinverse to obtain $S_{mn}$. 

\begin{figure} 
\includegraphics[width=0.8\textwidth]{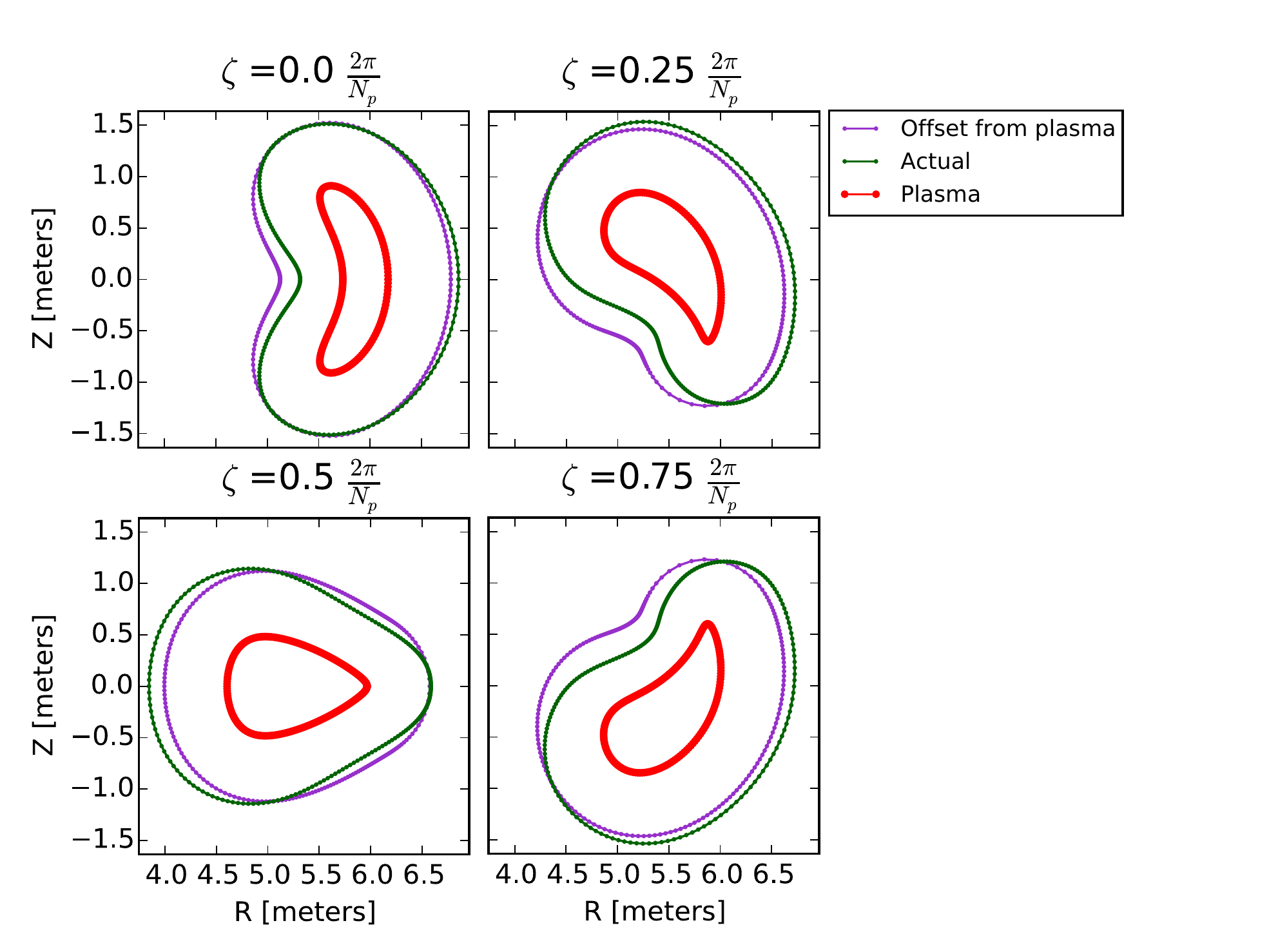}
\caption{The cross sections of the two winding surfaces used to compute $S_{\chi^2_B}$ and $S_{\norm{\bm{K}}_2}$ are shown in the poloidal plane.}
\label{w7x3surf}
\end{figure}

\begin{figure}
\includegraphics[width=.8\textwidth]{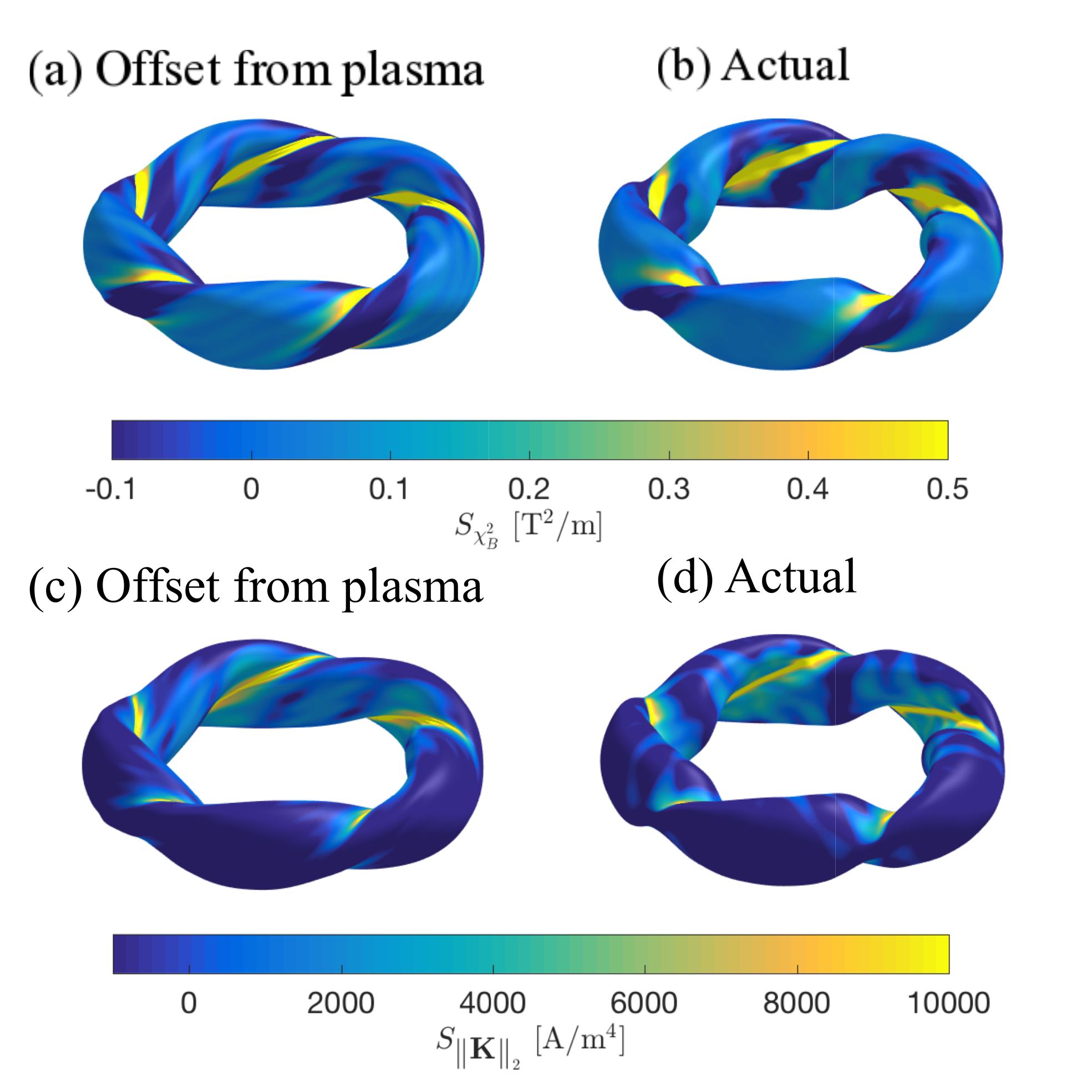}
\caption{Surface sensitivity functions for $\chi^2_K$ (upper subplots) and $\chi^2_B$ (lower subplots). These functions are computed using the W7-X plasma surface and a uniform offset winding surface from the plasma surface with $d_{\text{coil-plasma}} = 0.61$ m ((a) and (c)) and the actual winding surface ((b) and (d)). The region of increased $S_{\chi^2_B}$ corresponds with concave regions of the plasma surface (see figure \ref{dminandp2}(b)). Regions of large positive $\norm{\bm{K}}_2$ correspond to regions with increased $K$ (see figure \ref{w7x_K})}
\label{w7x_S}
\end{figure}

We compute $S_{\chi^2_B}$ and $S_{\norm{\bm{K}}_2}$ (figure \ref{w7x_S}) at fixed $\lambda$. These quantities are computed on the actual W7-X winding surface and a surface uniformly offset from the plasma surface with $d_{\text{coil-plasma}} = 0.61$ m (the area-averaged $d_{\text{coil-plasma}}$ over the actual surface). We consider surfaces that are equidistant from the plasma surface on average as $S$  scales inversely with $A_{\text{coil}}$. The poloidal cross-sections of these surfaces are shown in figure \ref{w7x3surf}. For each surface $\lambda$ is chosen to achieve $K_{\text{max}} = 7.7$ MA/m as was used in section \ref{w7x_results}. On both surfaces we observe a narrow region featuring a large positive $S_{\chi^2_B}$, indicating that $d_{\text{coil-plasma}}$ should decrease at that location in order that $\chi^2_B$ decreases. This corresponds to locations on the plasma surface with significant concavity (see figure \ref{dminandp2}(b)). 
The maximum $S_{\chi^2_B}$ occurs at $\zeta = 0.15 \, 2\pi/N_p$ on both surfaces (see figure \ref{w7x_surf}). 
In comparison with this region, the magnitude of $S_{\chi^2_B}$ is relatively small over the majority of the area of the surfaces shown, demonstrating that engineering tolerances might be more relaxed in these locations. There is also a region of negative $S_{\chi^2_B}$ near $\zeta = 0.5 \, 2\pi/N_p$ and $\theta =0$. This is the `tip' of the triangle-shaped cross-section, where $d_{\text{coil-plasma}}$ was increased over the course of the optimization (figures \ref{alpha2_scan}, \ref{alpha1_scan}, and \ref{w7x_surf}). We find that $S_{\chi^2_B}$ computed on the actual winding surface has similar trends to that computed on the surface uniformly offset from the plasma. Although on average these surfaces are equidistant from the plasma surface, the magnitude of $S_{\chi^2_B}$ is higher on the actual winding surface over much of the area. This indicates that the surface sensitivity function depends on the specific geometry of the winding surface. We have computed $S_{\chi^2_B}$ for several other winding surfaces with varying $d_{\text{coil-plasma}}$. Regardless of the winding surface chosen, we observe increased sensitivity in the concave regions. 

\begin{figure}
\includegraphics[width=0.8\textwidth]{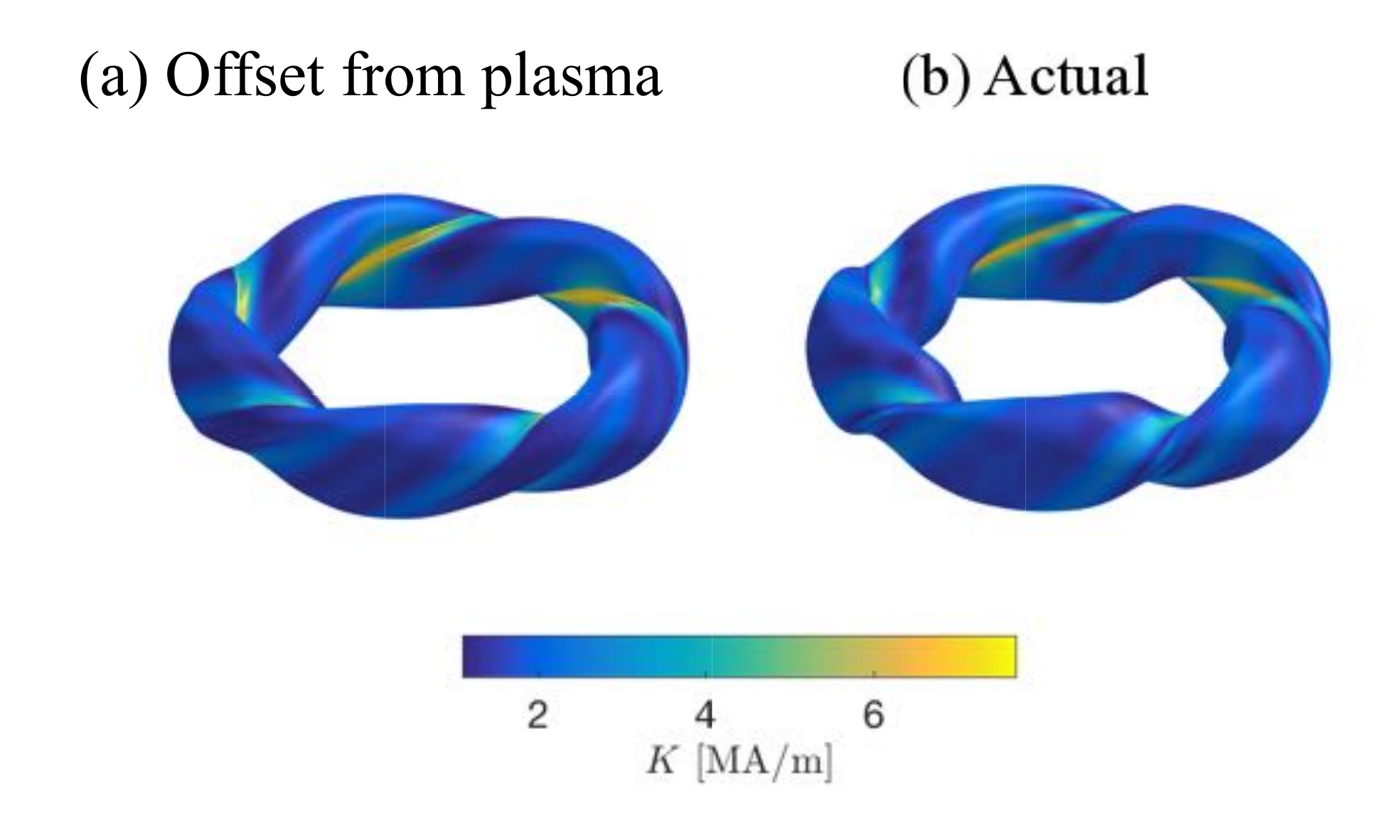}
\caption{Current density magnitude, $K$, computed from \texttt{REGCOIL} using the W7-X plasma surface and (a) a uniform offset winding surface from the plasma surface with $d_{\text{coil-plasma}} = 0.61$ m and (b) the actual winding surface.}
\label{w7x_K}
\end{figure}

The quantity $S_{\norm{\bm{K}}_2}$ roughly quantifies how coil complexity changes with normal displacements of the coil surface. In view of figure \ref{w7x_K}, the locations of large $S_{\norm{\bm{K}}_2}$ overlap with areas of increased $K$. On the actual winding surface, the maximum of $S_{\norm{\bm{K}}_2}$ occurs near the location of closest approach between coils (two rightmost coils in figure \ref{w7x_coils}(a)). The sensitivity functions $S_{\norm{\bm{K}}_2}$ and $S_{\chi^2_B}$ have very similar trends. The concave regions of the plasma surface are difficult to produce with external coils, resulting in increased coil complexity and $K$. Therefore, $\norm{\bm{K}}_2$ is most sensitive to displacements of the coil-winding surface in these regions. 

Studies of the plasma magnetic field sensitivity to perturbations of the coil placement on NCSX similarly found that coil errors on the inboard side in regions of small $d_{\text{coil-plasma}}$ had a significant effect on flux surface quality \cite{Williamson2005}. The necessity of small $d_{\text{coil-plasma}}$ for bean-shaped plasmas has been noted in many coil optimization efforts \cite{Strickler2002,Guebaly2008} and has been demonstrated by evaluating the singular value decomposition of the discretized Biot-Savart integral operator \cite{Landreman2016}. We are able to identify these regions where fidelity of the plasma surface requires tighter tolerance on coil positions using the surface sensitivity function. 

\section{Metrics for Configuration Optimization}
\label{sect_configopt}

Historically, stellarator design has proceeded by first optimizing an equilibrium based on various desired properties, such as neoclassical transport and MHD stability. Calculating the coils is a second step, done only after the equilibrium has been determined. The results presented here and in \cite{Landreman2016} indicate that the concave regions of the surface are both the areas where the optimizing routine chooses winding surfaces that lie close to the plasma and where the sensitivity to winding surface position is highest. 

The regions of concavity can be determined by considering the principal curvatures of the plasma surface. Let $\bm{n}$ represent the normal vector at the plasma surface at some point $r_0$, then let $A_n$ represent a plane that includes this normal vector. The intersection of the plane and the surface makes a curve $\bm{r}$, which has curvature $\kappa_0$ at the point $r_0$, as calculated from (\ref{curvature_of_curve}). Then the two principal curvatures $P_1$ and $P_2$ represent the maximum and minimum curvatures, $\kappa_0$, from all possible planes $A_n$. The signs of $P_1$ and $P_2$ depend on the convention chosen for the normal vector $\bm{n}$. We choose the convention such that convex curves, $\bm{r}$, have positive curvature and concave curves have negative curvatures. Therefore, minima of the second principal curvature, $P_2$, represent regions on the surface where the concavity is maximum.

\begin{figure}
\includegraphics[width=0.8\textwidth]{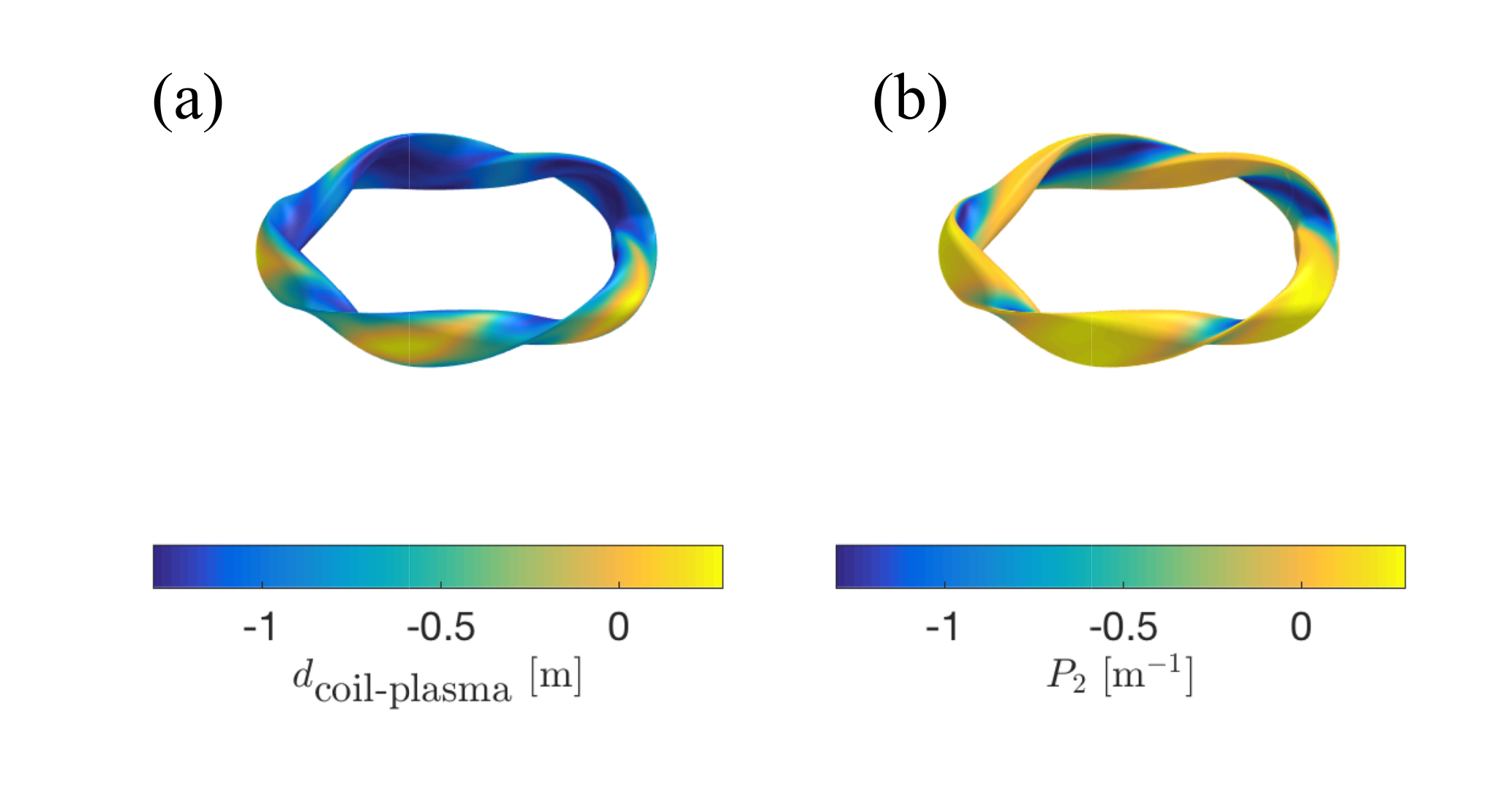}
\caption{(a) The minimum distance between the W7-X plasma surface and the optimized winding surface obtained in section \ref{w7x_results} and (b) the second principle curvature $P_2$ are shown as a function of location on the plasma surface. Locations of large negative $P_2$ coincide with regions where the optimization resulted in small $d_{\text{coil-plasma}}$.}
\label{dminandp2}
\end{figure}

The second principal curvature for the W7-X plasma is shown in figure \ref{dminandp2}(b). The regions of high concavity are represented by negative values of the second principal curvature. Although $P_2$ and the sensitivity functions are evaluated on different surfaces, we note that regions of high concavity (negative $P_2$) coincide with regions of high sensitivity (figure \ref{w7x_S}).
The regions of high concavity also correspond to the regions where the optimization procedure tends to place the winding surface closest to the plasma (see figure \ref{dminandp2}). We recognize that our winding surface optimization accounts for several engineering consideration in addition to reproducing the desired plasma surface. However, for a wide range of parameters the winding surfaces we obtain feature small $d_{\text{coil-plasma}}$ in the bean-shaped cross-sections (figures \ref{alpha2_scan} and \ref{alpha1_scan}). Thus $P_2$, which is exceedingly fast to compute, may serve as a target for optimization of the plasma configuration. By minimizing the regions of high concavity, it may be possible to find stellarator equilibria which are more amenable to coils that are positioned farther from the plasma. Any increase in the minimal distance between the plasma and the coils has implications for the size of a reactor, where the $d_{\text{coil-plasma}}$ is set by the blanket width.

\FloatBarrier
\section{Conclusions}
\label{sect_conclusions}

We have outlined a new method for optimization of the stellarator coil-winding surface using a continuous current potential approach. Rather than evolving filamentary coil shapes, we use \texttt{REGCOIL} to obtain the current density on a winding surface, and optimize the winding
surface using analytic gradients of the objective function. We have shown that we can indirectly improve the coil curvature and toroidal extent by targeting the root-mean-squared current density in our objective function (figure \ref{rmsKcoilcompare}). This approach offers several potential advantages over other nonlinear coil optimization tools.
\begin{enumerate}
\item The difficulty of the optimization is reduced by the application of the \texttt{REGCOIL} method, which takes the form of a linear least-squares system. The optimal coil shapes on a given winding surface can thus be efficiently and robustly computed. 
\item By fixing the maximum current density in order to obtain the regularization in \texttt{REGCOIL}, we eliminate the need to implement an additional equality constraint or arbitrary weight in the objective function. 
\item By using \texttt{REGCOIL} to compute coil shapes on a given surface, we are able to apply the adjoint method for computing derivatives (section \ref{sect_adjoint}). This allows us to reduce the number of function evaluations required during the nonlinear optimization by a factor of $\approx 50$. 
\item Given the critical role coil design plays in the stellarator optimization process, it is important to have many tools which approach the problem from different angles. Our approach differs from the other available nonlinear coil optimization applications \cite{Drevlak1998,Strickler2002,Strickler2004,Brown2015,Zhu2018} as we optimize a continuous current potential. 
\end{enumerate} 
We have demonstrated this method by optimizing coils for W7-X and HSX (sections \ref{w7x_results} and \ref{hsx_results}). We find that we are able to simultaneously decrease the integral-squared error in reproducing the plasma surface, increase the volume contained within the winding surface, maintain the minimum coil-plasma distance, and improve the coil metrics over \texttt{REGCOIL} solutions computed on the initial winding surfaces (tables \ref{table_w7x} and \ref{table_hsx}). Several features of these optimized winding surfaces are noteworthy. While the coil-plasma distance must be small in concave regions, it can increase greatly on the outboard, convex side of the bean cross-section. At triangle-shaped cross-sections, the winding surface obtains a somewhat `pinched' appearance (figures \ref{alpha1_scan}, \ref{w7x_surf}, and \ref{hsx_surf}). A similar W7-X winding surface shape has been obtained with the \texttt{ONSET} code (see ref. \cite{Drevlak1998}, figure 5). Further work is required to understand this behavior.

There are several limitations of this approach that should be noted. First, we have applied a local nonlinear optimization algorithm. This is a reasonable choice if the initial condition is close to a global optimum. We note that several global gradient-based optimization algorithms exist, which could be used if a global search is desired. Second, we currently have not added coil-specific metrics to our objective function (for example, curvature or length). This could be implemented if necessary for engineering purposes. 

We should also note that this application does not allow for the full benefits of adjoint methods. While adjoint methods significantly reduce CPU time if the solve is the computational bottleneck, this is not the case for the \texttt{REGCOIL} system, Other applications that are dominated by the linear solve CPU time would see increased benefits from the implementation of an adjoint method. In particular, the field of stellarator design could benefit from further incorporation of these methods in other aspects of the design process, such as computation of neoclassical transport and magnetic equilibria, as stellarators feature complex geometry with many free parameters describing a given configuration,

We demonstrate a technique for visualization of shape derivatives in real space rather than Fourier space. This surface sensitivity function describes how an objective function changes with respect to normal displacements of the winding surface. We apply this technique to visualize the derivatives of the integral-squared $B_n$ on the plasma surface and the root-mean-squared current density for the W7-X plasma surface and three winding surfaces (figure \ref{w7x_S}). This diagnostic identifies the concave regions as being very sensitive to the positions of coils, as has been observed from previous coil optimization efforts. This visualization technique could have potential applications for quantification of engineering tolerances in stellarator design and could be extended to the analysis of the sensitivity of other physics properties.


\appendix

\section{Adjoint derivative at fixed $K_{\text{max}}$}
\label{lambda_search}
We enforce $K_{\text{max}}=$ constant in the \texttt{REGCOIL} solve in order to obtain the regularization parameter $\lambda$ by requiring that the following constraint be satisfied within a given tolerance:
\begin{gather}
G(\Omega, \bm{\Phi}(\Omega,\lambda)) = K_{\text{max}}(\Omega, \bm{\Phi}(\Omega, \lambda)) - K^{\text{target}}_{\text{max}}  = 0 .
\label{K_constraint}
\end{gather}
Here $K^{\text{target}}_{\text{max}}$ is the target maximum current density and $\bm{\Phi}$ is chosen to satisfy the forward equation (\ref{forward}),
\begin{gather}
\bm{F} (\Omega,\bm{\Phi},\lambda) = \bm{A}(\Omega,\lambda) \bm{\Phi} - \bm{b}(\Omega,\lambda) = 0.
\label{forwardconstraint}
\end{gather}
A log-sum-exponent function is used to approximate the maximum function, similar to that used to approximate $d_{\text{coil-plasma}}$ (\ref{lse_d}).
\begin{gather}
K_{\text{max}} \approx K_{\text{max},\, \text{lse}} =  \frac{1}{p} \log \left( \frac{\int_{\text{coil}} d^2 A \,  \exp\left(p K\right)}{ A_{\text{coil}} } \right)
\end{gather}
We compute the total differential of $\bm{F}$,
\begin{gather}
d\bm{F} = \sum_j \left( \partder{\bm{A}}{\Omega_j} \bm{\Phi} - \partder{\bm{b}}{\Omega_j} \right) d\Omega_j + \bm{A} d \bm{\Phi} + \left(\bm{A}^K \bm{\Phi} - \bm{b}^K\right) d\lambda = 0.
\end{gather}
Here $\bm{A}^K = \partial \bm{A}/\partial \lambda$ and $\bm{b}^K = \partial \bm{b}/\partial \lambda$. We left multiply by $\bm{A}^{-1}$ and solve for $d\bm{\Phi}$.
\begin{gather}
d\bm{\Phi} = - \sum_j \bm{A}^{-1} \left( \partder{\bm{A}}{\Omega_j} \bm{\Phi} - \partder{\bm{b}}{\Omega_j} \right) d \Omega_j - \bm{A}^{-1} \left( \bm{A}^K \bm{\Phi} - \bm{b}^K \right) d \lambda
\label{dPhi}
\end{gather}
We also compute the total differential of $G$,
\begin{gather}
dG = \sum_j \partder{G}{\Omega_j} d\Omega_j + \partder{G}{\bm{\Phi}} \cdot d \bm{\Phi} = 0.
\end{gather}
Using the form for $d\bm{\Phi}$ (\ref{dPhi}), we compute $d\lambda$ in terms of $d\Omega_j$,
\begin{gather}
d\lambda = \left( \partder{G}{\bm{\Phi}} \cdot \left[ \bm{A}^{-1} \left( \bm{A}^K \bm{\Phi} - \bm{b}^K \right) \right] \right)^{-1} \sum_j \left( \partder{G}{\Omega_j} - \partder{G}{\bm{\Phi}}\cdot \left[ \bm{A}^{-1} \left( \partder{\bm{A}}{\Omega_j} \bm{\Phi} - \partder{\bm{b}}{\Omega_j} \right) \right] \right) d \Omega_j.
\label{dlambda}
\end{gather}
Using (\ref{dPhi}) and (\ref{dlambda}), the derivative of $\bm{\Phi}$ with respect to $\Omega_j$ subject to equations (\ref{K_constraint}) and (\ref{forwardconstraint}) is given by the following expression: 
\begin{multline}
\partder{\bm{\Phi}}{\Omega_j} \bigg \rvert_{\bm{F}=0, \, G = 0} = - \bm{A}^{-1} \left( \partder{\bm{A}}{\Omega_j} \bm{\Phi} - \partder{\bm{b}}{\Omega_j} \right)\\ - \frac{\bm{A}^{-1} \left( \bm{A}^K \bm{\Phi} - \bm{b}^K \right) }{ \partder{G}{\bm{\Phi}} \cdot \left[ \bm{A}^{-1} \left( \bm{A}^K \bm{\Phi} - \bm{b}^K \right) \right] } \left( \partder{G}{\Omega_j} - \partder{G}{\bm{\Phi}} \cdot \left[ \bm{A}^{-1} \left( \partder{\bm{A}}{\Omega_j} \bm{\Phi} - \partder{\bm{b}}{\Omega_j} \right) \right] \right).
\end{multline}
We use the adjoint method to avoid inverting the operator $\bm{A}$ for each $\Omega_j$,
\begin{multline}
\partder{\bm{\Phi}}{\Omega_j} \bigg \rvert_{\bm{F} = 0, \, G = 0} = - \bm{A}^{-1} \left( \partder{\bm{A}}{\Omega_j} \bm{\Phi} - \partder{\bm{b}}{\Omega_j} \right)\\ - \frac{\bm{A}^{-1} \left( \bm{A}^K \bm{\Phi} - \bm{b}^K \right) }{\partder{G}{\bm{\Phi}} \cdot \left[ \bm{A}^{-1} \left( \bm{A}^K \bm{\Phi} - \bm{b}^K \right) \right]} \left( \partder{G}{\Omega_j} - \left[ \left( \bm{A}^T \right)^{-1} \partder{G}{\bm{\Phi}} \right]  \cdot \left( \partder{\bm{A}}{\Omega_j} \bm{\Phi} - \partder{\bm{b}}{\Omega_j} \right) \right). 
\label{withadjoint}
\end{multline}
We introduce a new adjoint vector $\widetilde{\bm{q}}$,
defined to be the solution of
\begin{gather}
\bm{A}^T \widetilde{\bm{q}} = \partder{G}{\bm{\Phi}}.
\label{adjoint_2}
\end{gather}
Equation (\ref{withadjoint}) is then used to compute the derivatives of $\chi^2_B$ with respect to $\Omega_j$:
\begin{gather}
\partder{\chi^2_B}{\Omega_j} \bigg \rvert_{\bm{F} = 0, \, G = 0} = \partder{\chi^2_B}{\Omega_j} \bigg \rvert_{\bm{\Phi}, \lambda} + \partder{\chi^2_B}{\bm{\Phi}} \cdot \partder{\bm{\Phi}}{\Omega_j} \bigg \rvert_{\bm{F} = 0, \, G = 0}.
\end{gather}
This result can be written in terms of both adjoint variables, $\bm{q}$ and $\widetilde{\bm{q}}$:
\begin{multline}
\partder{\chi^2_B}{\Omega_j} \bigg \rvert_{\bm{F} = 0, \, G = 0} = \partder{\chi^2_B}{\Omega_j} \bigg \rvert_{\bm{\Phi}, \lambda} - \bm{q} \cdot \left( \partder{\bm{A}}{\Omega_j} \bm{\Phi} - \partder{\bm{b}}{\Omega_j} \right) - \frac{ \bm{q} \cdot \left(\bm{A}^K \bm{\Phi} - \bm{b}^K \right)}{  \widetilde{\bm{q}} \cdot \left( \bm{A}^K \bm{\Phi} - \bm{b}^K \right) } \left( \partder{G}{\Omega_j} - \widetilde{\bm{q}} \cdot \left( \partder{\bm{A}}{\Omega_j} \bm{\Phi} - \partder{\bm{b}}{\Omega_j} \right) \right).
\end{multline}
The same method is used to compute derivatives of $\norm{\bm{K}}_2$. So, to obtain the derivatives at fixed $K_{\text{max}}$, we compute a solution to the two adjoint equations, (\ref{adjoint}) and (\ref{adjoint_2}), in addition to the forward equation, (\ref{forward}).

\section*{Acknowledgements}
The authors would like to thank I. Abel and T. Antonsen for helpful input and discussions. This work was supported by the US Department of Energy through grants DE-FG02-93ER-54197 and DE-FC02-08ER-54964. The computations presented in this paper have used resources at the National Energy Research Scientific Computing Center (NERSC). 



\bibliographystyle{apsrev4-1}
\bibliography{AdjointRegcoil.bib}

\begin{thebibliography}{39}%
\makeatletter
\providecommand \@ifxundefined [1]{%
 \@ifx{#1\undefined}
}%
\providecommand \@ifnum [1]{%
 \ifnum #1\expandafter \@firstoftwo
 \else \expandafter \@secondoftwo
 \fi
}%
\providecommand \@ifx [1]{%
 \ifx #1\expandafter \@firstoftwo
 \else \expandafter \@secondoftwo
 \fi
}%
\providecommand \natexlab [1]{#1}%
\providecommand \enquote  [1]{``#1''}%
\providecommand \bibnamefont  [1]{#1}%
\providecommand \bibfnamefont [1]{#1}%
\providecommand \citenamefont [1]{#1}%
\providecommand \href@noop [0]{\@secondoftwo}%
\providecommand \href [0]{\begingroup \@sanitize@url \@href}%
\providecommand \@href[1]{\@@startlink{#1}\@@href}%
\providecommand \@@href[1]{\endgroup#1\@@endlink}%
\providecommand \@sanitize@url [0]{\catcode `\\12\catcode `\$12\catcode
  `\&12\catcode `\#12\catcode `\^12\catcode `\_12\catcode `\%12\relax}%
\providecommand \@@startlink[1]{}%
\providecommand \@@endlink[0]{}%
\providecommand \url  [0]{\begingroup\@sanitize@url \@url }%
\providecommand \@url [1]{\endgroup\@href {#1}{\urlprefix }}%
\providecommand \urlprefix  [0]{URL }%
\providecommand \Eprint [0]{\href }%
\providecommand \doibase [0]{http://dx.doi.org/}%
\providecommand \selectlanguage [0]{\@gobble}%
\providecommand \bibinfo  [0]{\@secondoftwo}%
\providecommand \bibfield  [0]{\@secondoftwo}%
\providecommand \translation [1]{[#1]}%
\providecommand \BibitemOpen [0]{}%
\providecommand \bibitemStop [0]{}%
\providecommand \bibitemNoStop [0]{.\EOS\space}%
\providecommand \EOS [0]{\spacefactor3000\relax}%
\providecommand \BibitemShut  [1]{\csname bibitem#1\endcsname}%
\let\auto@bib@innerbib\@empty
\bibitem [{\citenamefont {N\"{u}hrenberg}\ and\ \citenamefont
  {Zille}(1988)}]{Nuhrenberg1988}%
  \BibitemOpen
  \bibfield  {author} {\bibinfo {author} {\bibfnamefont {J.}~\bibnamefont
  {N\"{u}hrenberg}}\ and\ \bibinfo {author} {\bibfnamefont {R.}~\bibnamefont
  {Zille}},\ }\href {https://doi.org/10.1016/0375-9601(88)90080-1} {\bibfield
  {journal} {\bibinfo  {journal} {Physics Letters A}\ }\textbf {\bibinfo
  {volume} {129}},\ \bibinfo {pages} {113} (\bibinfo {year}
  {1988})}\BibitemShut {NoStop}%
\bibitem [{\citenamefont {El-Guebaly}\ \emph {et~al.}(2008)\citenamefont
  {El-Guebaly}, \citenamefont {Wilson}, \citenamefont {Henderson},
  \citenamefont {Sawan}, \citenamefont {Sviatoslavsky}, \citenamefont
  {Slaybaugh}, \citenamefont {Kiedrowski}, \citenamefont {Ibrahim},
  \citenamefont {Martin}, \citenamefont {Raffray}, \citenamefont {Malang},
  \citenamefont {Lyon}, \citenamefont {Ku}, \citenamefont {Wang}, \citenamefont
  {Bromberg}, \citenamefont {Merrill}, \citenamefont {Waganer}, \citenamefont
  {Najmabadi},\ and\ \citenamefont {the Aries-CS~Team}}]{Guebaly2008}%
  \BibitemOpen
  \bibfield  {author} {\bibinfo {author} {\bibfnamefont {L.}~\bibnamefont
  {El-Guebaly}}, \bibinfo {author} {\bibfnamefont {P.}~\bibnamefont {Wilson}},
  \bibinfo {author} {\bibfnamefont {D.}~\bibnamefont {Henderson}}, \bibinfo
  {author} {\bibfnamefont {M.}~\bibnamefont {Sawan}}, \bibinfo {author}
  {\bibfnamefont {G.}~\bibnamefont {Sviatoslavsky}}, \bibinfo {author}
  {\bibfnamefont {R.}~\bibnamefont {Slaybaugh}}, \bibinfo {author}
  {\bibfnamefont {B.}~\bibnamefont {Kiedrowski}}, \bibinfo {author}
  {\bibfnamefont {A.}~\bibnamefont {Ibrahim}}, \bibinfo {author} {\bibfnamefont
  {C.}~\bibnamefont {Martin}}, \bibinfo {author} {\bibfnamefont
  {R.}~\bibnamefont {Raffray}}, \bibinfo {author} {\bibfnamefont
  {S.}~\bibnamefont {Malang}}, \bibinfo {author} {\bibfnamefont
  {J.}~\bibnamefont {Lyon}}, \bibinfo {author} {\bibfnamefont {L.~P.}\
  \bibnamefont {Ku}}, \bibinfo {author} {\bibfnamefont {X.}~\bibnamefont
  {Wang}}, \bibinfo {author} {\bibfnamefont {L.}~\bibnamefont {Bromberg}},
  \bibinfo {author} {\bibfnamefont {B.}~\bibnamefont {Merrill}}, \bibinfo
  {author} {\bibfnamefont {L.}~\bibnamefont {Waganer}}, \bibinfo {author}
  {\bibfnamefont {F.}~\bibnamefont {Najmabadi}}, \ and\ \bibinfo {author}
  {\bibnamefont {the Aries-CS~Team}},\ }\href {\doibase 10.13182/FST54-747}
  {\bibfield  {journal} {\bibinfo  {journal} {Fusion Science and Technology}\
  }\textbf {\bibinfo {volume} {54}},\ \bibinfo {pages} {747} (\bibinfo {year}
  {2008})}\BibitemShut {NoStop}%
\bibitem [{\citenamefont {Landreman}\ and\ \citenamefont
  {Boozer}(2016)}]{Landreman2016}%
  \BibitemOpen
  \bibfield  {author} {\bibinfo {author} {\bibfnamefont {M.}~\bibnamefont
  {Landreman}}\ and\ \bibinfo {author} {\bibfnamefont {A.~H.}\ \bibnamefont
  {Boozer}},\ }\href {https://doi.org/10.1063/1.4943201} {\bibfield  {journal}
  {\bibinfo  {journal} {Physics of Plasmas}\ }\textbf {\bibinfo {volume}
  {23}},\ \bibinfo {pages} {032506} (\bibinfo {year} {2016})}\BibitemShut
  {NoStop}%
\bibitem [{\citenamefont {Merkel}(1987)}]{Merkel1987}%
  \BibitemOpen
  \bibfield  {author} {\bibinfo {author} {\bibfnamefont {P.}~\bibnamefont
  {Merkel}},\ }\href {\doibase 10.1088/0029-5515/27/5/018} {\bibfield
  {journal} {\bibinfo  {journal} {Nuclear Fusion}\ }\textbf {\bibinfo {volume}
  {27}},\ \bibinfo {pages} {867} (\bibinfo {year} {1987})}\BibitemShut
  {NoStop}%
\bibitem [{\citenamefont {Spong}\ and\ \citenamefont
  {Harris}(2010)}]{Spong2010}%
  \BibitemOpen
  \bibfield  {author} {\bibinfo {author} {\bibfnamefont {D.~A.}\ \bibnamefont
  {Spong}}\ and\ \bibinfo {author} {\bibfnamefont {J.~H.}\ \bibnamefont
  {Harris}},\ }\href {\doibase 10.1585/pfr.5.S2039} {\bibfield  {journal}
  {\bibinfo  {journal} {Plasma and Fusion Research}\ }\textbf {\bibinfo
  {volume} {5}},\ \bibinfo {pages} {S2039} (\bibinfo {year}
  {2010})}\BibitemShut {NoStop}%
\bibitem [{\citenamefont {Ku}\ and\ \citenamefont {Boozer}(2011)}]{Ku2011}%
  \BibitemOpen
  \bibfield  {author} {\bibinfo {author} {\bibfnamefont {L.~P.}\ \bibnamefont
  {Ku}}\ and\ \bibinfo {author} {\bibfnamefont {A.~H.}\ \bibnamefont
  {Boozer}},\ }\href {\doibase 10.1088/0029-5515/51/1/013004} {\bibfield
  {journal} {\bibinfo  {journal} {Nuclear Fusion}\ }\textbf {\bibinfo {volume}
  {51}},\ \bibinfo {pages} {013004} (\bibinfo {year} {2011})}\BibitemShut
  {NoStop}%
\bibitem [{\citenamefont {Drevlak}\ \emph {et~al.}(2013)\citenamefont
  {Drevlak}, \citenamefont {Brochard}, \citenamefont {Helander}, \citenamefont
  {Kisslinger}, \citenamefont {Mikhailov}, \citenamefont {N\"{u}hrenberg},
  \citenamefont {N\"{u}hrenberg},\ and\ \citenamefont {Turkin}}]{Drevlak2013}%
  \BibitemOpen
  \bibfield  {author} {\bibinfo {author} {\bibfnamefont {M.}~\bibnamefont
  {Drevlak}}, \bibinfo {author} {\bibfnamefont {F.}~\bibnamefont {Brochard}},
  \bibinfo {author} {\bibfnamefont {P.}~\bibnamefont {Helander}}, \bibinfo
  {author} {\bibfnamefont {J.}~\bibnamefont {Kisslinger}}, \bibinfo {author}
  {\bibfnamefont {M.}~\bibnamefont {Mikhailov}}, \bibinfo {author}
  {\bibfnamefont {C.}~\bibnamefont {N\"{u}hrenberg}}, \bibinfo {author}
  {\bibfnamefont {J.}~\bibnamefont {N\"{u}hrenberg}}, \ and\ \bibinfo {author}
  {\bibfnamefont {Y.}~\bibnamefont {Turkin}},\ }\href {\doibase
  10.1002/ctpp.201200055} {\bibfield  {journal} {\bibinfo  {journal}
  {Contributions to Plasma Physics}\ }\textbf {\bibinfo {volume} {53}},\
  \bibinfo {pages} {459} (\bibinfo {year} {2013})}\BibitemShut {NoStop}%
\bibitem [{\citenamefont {Pomphrey}\ \emph {et~al.}(2001)\citenamefont
  {Pomphrey}, \citenamefont {Berry}, \citenamefont {Boozer}, \citenamefont
  {Brooks}, \citenamefont {Hatcher}, \citenamefont {Hirshman}, \citenamefont
  {Ku}, \citenamefont {Miner}, \citenamefont {Mynick}, \citenamefont
  {Reiersen}, \citenamefont {Strickler},\ and\ \citenamefont
  {Valanju}}]{Pomphrey2001}%
  \BibitemOpen
  \bibfield  {author} {\bibinfo {author} {\bibfnamefont {N.}~\bibnamefont
  {Pomphrey}}, \bibinfo {author} {\bibfnamefont {L.}~\bibnamefont {Berry}},
  \bibinfo {author} {\bibfnamefont {A.}~\bibnamefont {Boozer}}, \bibinfo
  {author} {\bibfnamefont {A.}~\bibnamefont {Brooks}}, \bibinfo {author}
  {\bibfnamefont {R.}~\bibnamefont {Hatcher}}, \bibinfo {author} {\bibfnamefont
  {S.}~\bibnamefont {Hirshman}}, \bibinfo {author} {\bibfnamefont {L.-P.}\
  \bibnamefont {Ku}}, \bibinfo {author} {\bibfnamefont {W.}~\bibnamefont
  {Miner}}, \bibinfo {author} {\bibfnamefont {H.}~\bibnamefont {Mynick}},
  \bibinfo {author} {\bibfnamefont {W.}~\bibnamefont {Reiersen}}, \bibinfo
  {author} {\bibfnamefont {D.}~\bibnamefont {Strickler}}, \ and\ \bibinfo
  {author} {\bibfnamefont {P.}~\bibnamefont {Valanju}},\ }\href {\doibase
  10.1088/0029-5515/41/3/312} {\bibfield  {journal} {\bibinfo  {journal}
  {Nuclear Fusion}\ }\textbf {\bibinfo {volume} {41}},\ \bibinfo {pages} {339}
  (\bibinfo {year} {2001})}\BibitemShut {NoStop}%
\bibitem [{\citenamefont {Beidler}\ \emph {et~al.}(1990)\citenamefont
  {Beidler}, \citenamefont {Grieger}, \citenamefont {Herrnegger}, \citenamefont
  {Harmeyer}, \citenamefont {Kisslinger}, , \citenamefont {Lotz}, \citenamefont
  {Maassberg}, \citenamefont {Merkel}, \citenamefont {N{\"{u}}hrenberg},
  \citenamefont {Rau}, \citenamefont {Sapper}, \citenamefont {Sardei},
  \citenamefont {Scardovelli}, \citenamefont {Schl{\"{u}}ter},\ and\
  \citenamefont {Wobig}}]{Beidler1990}%
  \BibitemOpen
  \bibfield  {author} {\bibinfo {author} {\bibfnamefont {C.}~\bibnamefont
  {Beidler}}, \bibinfo {author} {\bibfnamefont {G.}~\bibnamefont {Grieger}},
  \bibinfo {author} {\bibfnamefont {F.}~\bibnamefont {Herrnegger}}, \bibinfo
  {author} {\bibfnamefont {E.}~\bibnamefont {Harmeyer}}, \bibinfo {author}
  {\bibfnamefont {J.}~\bibnamefont {Kisslinger}}, , \bibinfo {author}
  {\bibfnamefont {W.}~\bibnamefont {Lotz}}, \bibinfo {author} {\bibfnamefont
  {H.}~\bibnamefont {Maassberg}}, \bibinfo {author} {\bibfnamefont
  {P.}~\bibnamefont {Merkel}}, \bibinfo {author} {\bibfnamefont
  {J.}~\bibnamefont {N{\"{u}}hrenberg}}, \bibinfo {author} {\bibfnamefont
  {F.}~\bibnamefont {Rau}}, \bibinfo {author} {\bibfnamefont {J.}~\bibnamefont
  {Sapper}}, \bibinfo {author} {\bibfnamefont {F.}~\bibnamefont {Sardei}},
  \bibinfo {author} {\bibfnamefont {R.}~\bibnamefont {Scardovelli}}, \bibinfo
  {author} {\bibfnamefont {A.}~\bibnamefont {Schl{\"{u}}ter}}, \ and\ \bibinfo
  {author} {\bibfnamefont {H.}~\bibnamefont {Wobig}},\ }\href {\doibase
  10.13182/FST90-A29178} {\bibfield  {journal} {\bibinfo  {journal} {Fusion
  Technology}\ }\textbf {\bibinfo {volume} {17}},\ \bibinfo {pages} {148}
  (\bibinfo {year} {1990})}\BibitemShut {NoStop}%
\bibitem [{\citenamefont {Landreman}(2017)}]{Landreman2017}%
  \BibitemOpen
  \bibfield  {author} {\bibinfo {author} {\bibfnamefont {M.}~\bibnamefont
  {Landreman}},\ }\href {\doibase 10.1088/1741-4326/aa57d4} {\bibfield
  {journal} {\bibinfo  {journal} {Nuclear Fusion}\ }\textbf {\bibinfo {volume}
  {57}},\ \bibinfo {pages} {046003} (\bibinfo {year} {2017})}\BibitemShut
  {NoStop}%
\bibitem [{\citenamefont {{M. Drevlak}}(1998)}]{Drevlak1998}%
  \BibitemOpen
  \bibfield  {author} {\bibinfo {author} {\bibnamefont {{M. Drevlak}}},\ }\href
  {\doibase 10.13182/FST98-A21} {\bibfield  {journal} {\bibinfo  {journal}
  {Fusion Technol.}\ }\textbf {\bibinfo {volume} {33}},\ \bibinfo {pages} {106}
  (\bibinfo {year} {1998})}\BibitemShut {NoStop}%
\bibitem [{\citenamefont {Strickler}\ \emph {et~al.}(2002)\citenamefont
  {Strickler}, \citenamefont {Berry},\ and\ \citenamefont
  {Hirshman}}]{Strickler2002}%
  \BibitemOpen
  \bibfield  {author} {\bibinfo {author} {\bibfnamefont {D.~J.}\ \bibnamefont
  {Strickler}}, \bibinfo {author} {\bibfnamefont {L.~A.}\ \bibnamefont
  {Berry}}, \ and\ \bibinfo {author} {\bibfnamefont {S.~P.}\ \bibnamefont
  {Hirshman}},\ }\href {\doibase 10.13182/FST02-A206} {\bibfield  {journal}
  {\bibinfo  {journal} {Fusion Science and Technology}\ }\textbf {\bibinfo
  {volume} {41}},\ \bibinfo {pages} {107} (\bibinfo {year} {2002})}\BibitemShut
  {NoStop}%
\bibitem [{\citenamefont {Strickler}\ \emph {et~al.}(2004)\citenamefont
  {Strickler}, \citenamefont {Hirshman}, \citenamefont {Spong}, \citenamefont
  {Cole}, \citenamefont {Lyon}, \citenamefont {Nelson}, \citenamefont
  {Williamson},\ and\ \citenamefont {Ware}}]{Strickler2004}%
  \BibitemOpen
  \bibfield  {author} {\bibinfo {author} {\bibfnamefont {D.~J.}\ \bibnamefont
  {Strickler}}, \bibinfo {author} {\bibfnamefont {S.~P.}\ \bibnamefont
  {Hirshman}}, \bibinfo {author} {\bibfnamefont {D.~A.}\ \bibnamefont {Spong}},
  \bibinfo {author} {\bibfnamefont {M.~J.}\ \bibnamefont {Cole}}, \bibinfo
  {author} {\bibfnamefont {J.~F.}\ \bibnamefont {Lyon}}, \bibinfo {author}
  {\bibfnamefont {B.~E.}\ \bibnamefont {Nelson}}, \bibinfo {author}
  {\bibfnamefont {D.~E.}\ \bibnamefont {Williamson}}, \ and\ \bibinfo {author}
  {\bibfnamefont {A.~S.}\ \bibnamefont {Ware}},\ }\href {\doibase
  10.13182/FST04-A421} {\bibfield  {journal} {\bibinfo  {journal} {Fusion
  Science and Technology}\ }\textbf {\bibinfo {volume} {45}},\ \bibinfo {pages}
  {15} (\bibinfo {year} {2004})}\BibitemShut {NoStop}%
\bibitem [{\citenamefont {Zarnstorff}\ \emph {et~al.}(2001)\citenamefont
  {Zarnstorff}, \citenamefont {Berry}, \citenamefont {Brooks}, \citenamefont
  {Fredrickson}, \citenamefont {Fu}, \citenamefont {Hirshman}, \citenamefont
  {Hudson}, \citenamefont {Ku}, \citenamefont {Lazarus}, \citenamefont
  {Mikkelsen}, \citenamefont {Monticello}, \citenamefont {Neilson},
  \citenamefont {Pomphrey}, \citenamefont {Reiman}, \citenamefont {Spong},
  \citenamefont {Strickler}, \citenamefont {Boozer}, \citenamefont {Cooper},
  \citenamefont {Goldston}, \citenamefont {Hatcher}, \citenamefont {Isaev},
  \citenamefont {Kessel}, \citenamefont {Lewandowski}, \citenamefont {Lyon},
  \citenamefont {Merkel}, \citenamefont {Mynick}, \citenamefont {Nelson},
  \citenamefont {N\"{u}hrenberg}, \citenamefont {Redi}, \citenamefont
  {Reiersen}, \citenamefont {Rutherford}, \citenamefont {Sanchez},
  \citenamefont {Schmidt},\ and\ \citenamefont {White}}]{Zarnstorff2001}%
  \BibitemOpen
  \bibfield  {author} {\bibinfo {author} {\bibfnamefont {M.}~\bibnamefont
  {Zarnstorff}}, \bibinfo {author} {\bibfnamefont {L.}~\bibnamefont {Berry}},
  \bibinfo {author} {\bibfnamefont {A.}~\bibnamefont {Brooks}}, \bibinfo
  {author} {\bibfnamefont {E.}~\bibnamefont {Fredrickson}}, \bibinfo {author}
  {\bibfnamefont {G.-Y.}\ \bibnamefont {Fu}}, \bibinfo {author} {\bibfnamefont
  {S.}~\bibnamefont {Hirshman}}, \bibinfo {author} {\bibfnamefont
  {S.}~\bibnamefont {Hudson}}, \bibinfo {author} {\bibfnamefont {L.-P.}\
  \bibnamefont {Ku}}, \bibinfo {author} {\bibfnamefont {E.}~\bibnamefont
  {Lazarus}}, \bibinfo {author} {\bibfnamefont {D.}~\bibnamefont {Mikkelsen}},
  \bibinfo {author} {\bibfnamefont {D.}~\bibnamefont {Monticello}}, \bibinfo
  {author} {\bibfnamefont {G.~H.}\ \bibnamefont {Neilson}}, \bibinfo {author}
  {\bibfnamefont {N.}~\bibnamefont {Pomphrey}}, \bibinfo {author}
  {\bibfnamefont {A.}~\bibnamefont {Reiman}}, \bibinfo {author} {\bibfnamefont
  {D.}~\bibnamefont {Spong}}, \bibinfo {author} {\bibfnamefont
  {D.}~\bibnamefont {Strickler}}, \bibinfo {author} {\bibfnamefont
  {A.}~\bibnamefont {Boozer}}, \bibinfo {author} {\bibfnamefont {W.~A.}\
  \bibnamefont {Cooper}}, \bibinfo {author} {\bibfnamefont {R.~J.}\
  \bibnamefont {Goldston}}, \bibinfo {author} {\bibfnamefont {R.}~\bibnamefont
  {Hatcher}}, \bibinfo {author} {\bibfnamefont {M.~Y.}\ \bibnamefont {Isaev}},
  \bibinfo {author} {\bibfnamefont {C.}~\bibnamefont {Kessel}}, \bibinfo
  {author} {\bibfnamefont {J.}~\bibnamefont {Lewandowski}}, \bibinfo {author}
  {\bibfnamefont {J.}~\bibnamefont {Lyon}}, \bibinfo {author} {\bibfnamefont
  {P.}~\bibnamefont {Merkel}}, \bibinfo {author} {\bibfnamefont
  {H.}~\bibnamefont {Mynick}}, \bibinfo {author} {\bibfnamefont
  {B.}~\bibnamefont {Nelson}}, \bibinfo {author} {\bibfnamefont
  {C.}~\bibnamefont {N\"{u}hrenberg}}, \bibinfo {author} {\bibfnamefont
  {M.}~\bibnamefont {Redi}}, \bibinfo {author} {\bibfnamefont {W.}~\bibnamefont
  {Reiersen}}, \bibinfo {author} {\bibfnamefont {P.}~\bibnamefont
  {Rutherford}}, \bibinfo {author} {\bibfnamefont {R.}~\bibnamefont {Sanchez}},
  \bibinfo {author} {\bibfnamefont {J.}~\bibnamefont {Schmidt}}, \ and\
  \bibinfo {author} {\bibfnamefont {R.}~\bibnamefont {White}},\ }\href
  {\doibase 10.1088/0741-3335/43/12A/318} {\bibfield  {journal} {\bibinfo
  {journal} {Plasma Physics and Controlled Fusion}\ }\textbf {\bibinfo {volume}
  {43}},\ \bibinfo {pages} {A237} (\bibinfo {year} {2001})}\BibitemShut
  {NoStop}%
\bibitem [{\citenamefont {Brown}\ \emph {et~al.}(2015)\citenamefont {Brown},
  \citenamefont {Breslau}, \citenamefont {Gates}, \citenamefont {Pomphrey},\
  and\ \citenamefont {Zolfaghari}}]{Brown2015}%
  \BibitemOpen
  \bibfield  {author} {\bibinfo {author} {\bibfnamefont {T.}~\bibnamefont
  {Brown}}, \bibinfo {author} {\bibfnamefont {J.}~\bibnamefont {Breslau}},
  \bibinfo {author} {\bibfnamefont {D.}~\bibnamefont {Gates}}, \bibinfo
  {author} {\bibfnamefont {N.}~\bibnamefont {Pomphrey}}, \ and\ \bibinfo
  {author} {\bibfnamefont {A.}~\bibnamefont {Zolfaghari}},\ }in\ \href
  {\doibase 10.1109/SOFE.2015.7482426} {\emph {\bibinfo {booktitle} {IEEE 26th
  Symposium on Fusion Engineering (SOFE)}}}\ (\bibinfo {address} {Austin,
  Texas},\ \bibinfo {year} {2015})\BibitemShut {NoStop}%
\bibitem [{\citenamefont {Zhu}\ \emph {et~al.}(2018)\citenamefont {Zhu},
  \citenamefont {Hudson}, \citenamefont {Song},\ and\ \citenamefont
  {Wan}}]{Zhu2018}%
  \BibitemOpen
  \bibfield  {author} {\bibinfo {author} {\bibfnamefont {C.}~\bibnamefont
  {Zhu}}, \bibinfo {author} {\bibfnamefont {S.~R.}\ \bibnamefont {Hudson}},
  \bibinfo {author} {\bibfnamefont {Y.}~\bibnamefont {Song}}, \ and\ \bibinfo
  {author} {\bibfnamefont {Y.}~\bibnamefont {Wan}},\ }\href
  {https://doi.org/10.1088/1741-4326/aa8e0a} {\bibfield  {journal} {\bibinfo
  {journal} {Nuclear Fusion}\ }\textbf {\bibinfo {volume} {58}},\ \bibinfo
  {pages} {016008} (\bibinfo {year} {2018})}\BibitemShut {NoStop}%
\bibitem [{\citenamefont {Nocedal}\ and\ \citenamefont
  {Wright}(2006)}]{Nocedal2006}%
  \BibitemOpen
  \bibfield  {author} {\bibinfo {author} {\bibfnamefont {J.}~\bibnamefont
  {Nocedal}}\ and\ \bibinfo {author} {\bibfnamefont {S.~J.}\ \bibnamefont
  {Wright}},\ }\href@noop {} {\emph {\bibinfo {title} {Numerical
  Optimization}}}\ (\bibinfo  {publisher} {Springer},\ \bibinfo {year} {2006})\
  pp.\ \bibinfo {pages} {193--194}\BibitemShut {NoStop}%
\bibitem [{\citenamefont {Roy}\ and\ \citenamefont
  {Oberkampf}(2011)}]{Roy2011}%
  \BibitemOpen
  \bibfield  {author} {\bibinfo {author} {\bibfnamefont {C.~J.}\ \bibnamefont
  {Roy}}\ and\ \bibinfo {author} {\bibfnamefont {W.~L.}\ \bibnamefont
  {Oberkampf}},\ }\href {\doibase 10.1016/j.cma.2011.03.016} {\bibfield
  {journal} {\bibinfo  {journal} {Computer Methods in Applied Mechanics and
  Engineering}\ }\textbf {\bibinfo {volume} {200}},\ \bibinfo {pages} {2131}
  (\bibinfo {year} {2011})}\BibitemShut {NoStop}%
\bibitem [{\citenamefont {Othmer}(2008)}]{Othmer2008}%
  \BibitemOpen
  \bibfield  {author} {\bibinfo {author} {\bibfnamefont {C.}~\bibnamefont
  {Othmer}},\ }\href {\doibase 10.1002/fld.1770} {\bibfield  {journal}
  {\bibinfo  {journal} {International Journal for Numerical Methods in Fluids}\
  }\textbf {\bibinfo {volume} {58}},\ \bibinfo {pages} {861} (\bibinfo {year}
  {2008})}\BibitemShut {NoStop}%
\bibitem [{\citenamefont {Othmer}(2014)}]{Othmer2014}%
  \BibitemOpen
  \bibfield  {author} {\bibinfo {author} {\bibfnamefont {C.}~\bibnamefont
  {Othmer}},\ }\href {\doibase 10.1186/2190-5983-4-6} {\bibfield  {journal}
  {\bibinfo  {journal} {Journal of Mathematics in Industry}\ }\textbf {\bibinfo
  {volume} {4}},\ \bibinfo {pages} {6} (\bibinfo {year} {2014})}\BibitemShut
  {NoStop}%
\bibitem [{\citenamefont {Pironneau}(1974)}]{Pironneau1974}%
  \BibitemOpen
  \bibfield  {author} {\bibinfo {author} {\bibfnamefont {O.}~\bibnamefont
  {Pironneau}},\ }\href {https://doi.org/10.1017/S0022112074002023} {\bibfield
  {journal} {\bibinfo  {journal} {Journal of Fluid Mechanics}\ }\textbf
  {\bibinfo {volume} {64}},\ \bibinfo {pages} {97} (\bibinfo {year}
  {1974})}\BibitemShut {NoStop}%
\bibitem [{\citenamefont {Kuruvila}\ \emph {et~al.}(1995)\citenamefont
  {Kuruvila}, \citenamefont {Ta'asan},\ and\ \citenamefont
  {Salas}}]{Kuruvila1995}%
  \BibitemOpen
  \bibfield  {author} {\bibinfo {author} {\bibfnamefont {G.}~\bibnamefont
  {Kuruvila}}, \bibinfo {author} {\bibfnamefont {S.}~\bibnamefont {Ta'asan}}, \
  and\ \bibinfo {author} {\bibfnamefont {M.}~\bibnamefont {Salas}},\ }in\ \href
  {\doibase doi:10.2514/6.1995-478} {\emph {\bibinfo {booktitle} {33rd
  Aerospace Sciences Meeting and Exhibit}}}\ (\bibinfo {address} {Reno,
  Nevada},\ \bibinfo {year} {1995})\BibitemShut {NoStop}%
\bibitem [{\citenamefont {Jameson}\ \emph {et~al.}(1998)\citenamefont
  {Jameson}, \citenamefont {Martinelli},\ and\ \citenamefont
  {Pierce}}]{Jameson1998}%
  \BibitemOpen
  \bibfield  {author} {\bibinfo {author} {\bibfnamefont {A.}~\bibnamefont
  {Jameson}}, \bibinfo {author} {\bibfnamefont {L.}~\bibnamefont {Martinelli}},
  \ and\ \bibinfo {author} {\bibfnamefont {N.}~\bibnamefont {Pierce}},\ }\href
  {\doibase 10.1007/s001620050060} {\bibfield  {journal} {\bibinfo  {journal}
  {Theoretical and Computational Fluid Dynamics}\ }\textbf {\bibinfo {volume}
  {10}},\ \bibinfo {pages} {213} (\bibinfo {year} {1998})}\BibitemShut
  {NoStop}%
\bibitem [{\citenamefont {Anderson}\ and\ \citenamefont
  {Venkatakrishnan}(1999)}]{Anderson1999}%
  \BibitemOpen
  \bibfield  {author} {\bibinfo {author} {\bibfnamefont {W.~K.}\ \bibnamefont
  {Anderson}}\ and\ \bibinfo {author} {\bibfnamefont {V.}~\bibnamefont
  {Venkatakrishnan}},\ }\href {\doibase 10.1016/S0045-7930(98)00041-3}
  {\bibfield  {journal} {\bibinfo  {journal} {Computers and Fluids}\ }\textbf
  {\bibinfo {volume} {28}},\ \bibinfo {pages} {443} (\bibinfo {year}
  {1999})}\BibitemShut {NoStop}%
\bibitem [{\citenamefont {Kim}\ \emph {et~al.}(2001)\citenamefont {Kim},
  \citenamefont {Coster}, \citenamefont {Neuhauser}, \citenamefont
  {Schneider},\ and\ \citenamefont {the ASDEX-Upgrade~Team}}]{Kim2001}%
  \BibitemOpen
  \bibfield  {author} {\bibinfo {author} {\bibfnamefont {J.}~\bibnamefont
  {Kim}}, \bibinfo {author} {\bibfnamefont {D.~P.}\ \bibnamefont {Coster}},
  \bibinfo {author} {\bibfnamefont {J.}~\bibnamefont {Neuhauser}}, \bibinfo
  {author} {\bibfnamefont {R.}~\bibnamefont {Schneider}}, \ and\ \bibinfo
  {author} {\bibnamefont {the ASDEX-Upgrade~Team}},\ }\href {\doibase
  10.1016/S0022-3115(00)00599-7} {\bibfield  {journal} {\bibinfo  {journal}
  {Journal of Nuclear Materials}\ }\textbf {\bibinfo {volume} {293}},\ \bibinfo
  {pages} {644} (\bibinfo {year} {2001})}\BibitemShut {NoStop}%
\bibitem [{\citenamefont {Baelmans}\ \emph {et~al.}(2017)\citenamefont
  {Baelmans}, \citenamefont {Blommaert}, \citenamefont {Dekeyser},\ and\
  \citenamefont {{Van Oevelen}}}]{Baelmans2017}%
  \BibitemOpen
  \bibfield  {author} {\bibinfo {author} {\bibfnamefont {M.}~\bibnamefont
  {Baelmans}}, \bibinfo {author} {\bibfnamefont {M.}~\bibnamefont {Blommaert}},
  \bibinfo {author} {\bibfnamefont {W.}~\bibnamefont {Dekeyser}}, \ and\
  \bibinfo {author} {\bibfnamefont {T.}~\bibnamefont {{Van Oevelen}}},\ }\href
  {\doibase 10.1088/1741-4326/57/3/036022} {\bibfield  {journal} {\bibinfo
  {journal} {Nuclear Fusion}\ }\textbf {\bibinfo {volume} {57}},\ \bibinfo
  {pages} {036022} (\bibinfo {year} {2017})}\BibitemShut {NoStop}%
\bibitem [{\citenamefont {Jia}\ \emph {et~al.}(2014)\citenamefont {Jia},
  \citenamefont {Liu}, \citenamefont {Zaitsev}, \citenamefont {Hennig},\ and\
  \citenamefont {Korvink}}]{Jia2014}%
  \BibitemOpen
  \bibfield  {author} {\bibinfo {author} {\bibfnamefont {F.}~\bibnamefont
  {Jia}}, \bibinfo {author} {\bibfnamefont {Z.}~\bibnamefont {Liu}}, \bibinfo
  {author} {\bibfnamefont {M.}~\bibnamefont {Zaitsev}}, \bibinfo {author}
  {\bibfnamefont {J.}~\bibnamefont {Hennig}}, \ and\ \bibinfo {author}
  {\bibfnamefont {J.~G.}\ \bibnamefont {Korvink}},\ }\href {\doibase
  10.1007/s00158-013-0992-8} {\bibfield  {journal} {\bibinfo  {journal}
  {Structural and Multidisciplinary Optimization}\ }\textbf {\bibinfo {volume}
  {49}},\ \bibinfo {pages} {523} (\bibinfo {year} {2014})}\BibitemShut
  {NoStop}%
\bibitem [{\citenamefont {Turner}(1993)}]{Turner1993}%
  \BibitemOpen
  \bibfield  {author} {\bibinfo {author} {\bibfnamefont {R.}~\bibnamefont
  {Turner}},\ }\href {\doibase 10.1016/0730-725X(93)90209-V} {\bibfield
  {journal} {\bibinfo  {journal} {Magnetic Resonance Imaging}\ }\textbf
  {\bibinfo {volume} {11}},\ \bibinfo {pages} {903} (\bibinfo {year}
  {1993})}\BibitemShut {NoStop}%
\bibitem [{\citenamefont {Forbes}\ \emph {et~al.}(2005)\citenamefont {Forbes},
  \citenamefont {Brideson},\ and\ \citenamefont {Crozier}}]{Forbes2005}%
  \BibitemOpen
  \bibfield  {author} {\bibinfo {author} {\bibfnamefont {L.~K.}\ \bibnamefont
  {Forbes}}, \bibinfo {author} {\bibfnamefont {M.~A.}\ \bibnamefont
  {Brideson}}, \ and\ \bibinfo {author} {\bibfnamefont {S.}~\bibnamefont
  {Crozier}},\ }\href {\doibase 10.1109/TMAG.2005.847638} {\bibfield  {journal}
  {\bibinfo  {journal} {IEEE Transactions on Magnetics}\ }\textbf {\bibinfo
  {volume} {41}},\ \bibinfo {pages} {2134} (\bibinfo {year}
  {2005})}\BibitemShut {NoStop}%
\bibitem [{\citenamefont {Forbes}\ and\ \citenamefont
  {Crozier}(2001)}]{Forbes2001}%
  \BibitemOpen
  \bibfield  {author} {\bibinfo {author} {\bibfnamefont {L.~K.}\ \bibnamefont
  {Forbes}}\ and\ \bibinfo {author} {\bibfnamefont {S.}~\bibnamefont
  {Crozier}},\ }\href {\doibase 10.1088/0022-3727/34/24/305} {\bibfield
  {journal} {\bibinfo  {journal} {Journal of Physics D: Applied Physics}\
  }\textbf {\bibinfo {volume} {34}},\ \bibinfo {pages} {3447} (\bibinfo {year}
  {2001})}\BibitemShut {NoStop}%
\bibitem [{\citenamefont {Hirshman}\ and\ \citenamefont
  {Meier}(1985)}]{Hirshman1985}%
  \BibitemOpen
  \bibfield  {author} {\bibinfo {author} {\bibfnamefont {S.~P.}\ \bibnamefont
  {Hirshman}}\ and\ \bibinfo {author} {\bibfnamefont {H.~K.}\ \bibnamefont
  {Meier}},\ }\href {\doibase 10.1063/1.864972} {\bibfield  {journal} {\bibinfo
   {journal} {Physics of Fluids}\ }\textbf {\bibinfo {volume} {28}},\ \bibinfo
  {pages} {1387} (\bibinfo {year} {1985})}\BibitemShut {NoStop}%
\bibitem [{\citenamefont {Johnson}(2014)}]{NLOPT}%
  \BibitemOpen
  \bibfield  {author} {\bibinfo {author} {\bibfnamefont {S.~G.}\ \bibnamefont
  {Johnson}},\ }\href {http://ab-initio.mit.edu/nlopt} {\enquote {\bibinfo
  {title} {The {NLopt} nonlinear-optimization package},}\ } (\bibinfo {year}
  {2014})\BibitemShut {NoStop}%
\bibitem [{\citenamefont {Svanberg}(2002)}]{Svanberg2002}%
  \BibitemOpen
  \bibfield  {author} {\bibinfo {author} {\bibfnamefont {K.}~\bibnamefont
  {Svanberg}},\ }\href {\doibase 10.1137/S1052623499362822} {\bibfield
  {journal} {\bibinfo  {journal} {SIAM Journal of Optimization}\ }\textbf
  {\bibinfo {volume} {12}},\ \bibinfo {pages} {555} (\bibinfo {year}
  {2002})}\BibitemShut {NoStop}%
\bibitem [{\citenamefont {Boyd}\ and\ \citenamefont
  {Vandenberghe}(2004)}]{Boyd2004}%
  \BibitemOpen
  \bibfield  {author} {\bibinfo {author} {\bibfnamefont {S.}~\bibnamefont
  {Boyd}}\ and\ \bibinfo {author} {\bibfnamefont {L.}~\bibnamefont
  {Vandenberghe}},\ }\href@noop {} {\emph {\bibinfo {title} {Convex
  Optimization}}}\ (\bibinfo  {publisher} {Cambridge University Press},\
  \bibinfo {year} {2004})\ p.~\bibinfo {pages} {72}\BibitemShut {NoStop}%
\bibitem [{\citenamefont {Rust}\ \emph {et~al.}(2011)\citenamefont {Rust},
  \citenamefont {Heinemann}, \citenamefont {Mendelevitch}, \citenamefont
  {Peacock},\ and\ \citenamefont {Smirnow}}]{Rust2011}%
  \BibitemOpen
  \bibfield  {author} {\bibinfo {author} {\bibfnamefont {N.}~\bibnamefont
  {Rust}}, \bibinfo {author} {\bibfnamefont {B.}~\bibnamefont {Heinemann}},
  \bibinfo {author} {\bibfnamefont {B.}~\bibnamefont {Mendelevitch}}, \bibinfo
  {author} {\bibfnamefont {A.}~\bibnamefont {Peacock}}, \ and\ \bibinfo
  {author} {\bibfnamefont {M.}~\bibnamefont {Smirnow}},\ }\href {\doibase
  10.1016/j.fusengdes.2011.03.054} {\bibfield  {journal} {\bibinfo  {journal}
  {Fusion Engineering and Design}\ }\textbf {\bibinfo {volume} {86}},\ \bibinfo
  {pages} {728} (\bibinfo {year} {2011})}\BibitemShut {NoStop}%
\bibitem [{\citenamefont {Delfour}\ and\ \citenamefont
  {Zol\'{e}sio}(2011)}]{Delfour2011}%
  \BibitemOpen
  \bibfield  {author} {\bibinfo {author} {\bibfnamefont {M.~C.}\ \bibnamefont
  {Delfour}}\ and\ \bibinfo {author} {\bibfnamefont {J.-P.}\ \bibnamefont
  {Zol\'{e}sio}},\ }\href@noop {} {\emph {\bibinfo {title} {Shapes and
  Geometries}}}\ (\bibinfo  {publisher} {Society for Industrial and Applied
  Mathematics},\ \bibinfo {year} {2011})\ pp.\ \bibinfo {pages}
  {480--481}\BibitemShut {NoStop}%
\bibitem [{\citenamefont {Novotny}\ and\ \citenamefont
  {Sokolowski}(2013)}]{Novotny2013}%
  \BibitemOpen
  \bibfield  {author} {\bibinfo {author} {\bibfnamefont {A.~A.}\ \bibnamefont
  {Novotny}}\ and\ \bibinfo {author} {\bibfnamefont {J.}~\bibnamefont
  {Sokolowski}},\ }\href@noop {} {\emph {\bibinfo {title} {Topological
  Derivatives in Shape Optimization}}}\ (\bibinfo  {publisher} {Springer},\
  \bibinfo {year} {2013})\ pp.\ \bibinfo {pages} {35--38}\BibitemShut {NoStop}%
\bibitem [{\citenamefont {Landreman}(tion)}]{Landreman2018}%
  \BibitemOpen
  \bibfield  {author} {\bibinfo {author} {\bibfnamefont {M.}~\bibnamefont
  {Landreman}},\ }\href@noop {} {\  (\bibinfo {year} {in
  preparation})}\BibitemShut {NoStop}%
\bibitem [{\citenamefont {Williamson}\ \emph {et~al.}(2005)\citenamefont
  {Williamson}, \citenamefont {Brooks}, \citenamefont {Brown}, \citenamefont
  {Chrzanowski}, \citenamefont {Cole}, \citenamefont {Fan}, \citenamefont
  {Freudenberg}, \citenamefont {Fogarty}, \citenamefont {Hargrove},
  \citenamefont {Heitzenroeder}, \citenamefont {Lovett}, \citenamefont
  {Miller}, \citenamefont {Myatt}, \citenamefont {Nelson}, \citenamefont
  {Reiersen},\ and\ \citenamefont {Strickler}}]{Williamson2005}%
  \BibitemOpen
  \bibfield  {author} {\bibinfo {author} {\bibfnamefont {D.}~\bibnamefont
  {Williamson}}, \bibinfo {author} {\bibfnamefont {A.}~\bibnamefont {Brooks}},
  \bibinfo {author} {\bibfnamefont {T.}~\bibnamefont {Brown}}, \bibinfo
  {author} {\bibfnamefont {J.}~\bibnamefont {Chrzanowski}}, \bibinfo {author}
  {\bibfnamefont {M.}~\bibnamefont {Cole}}, \bibinfo {author} {\bibfnamefont
  {H.-M.}\ \bibnamefont {Fan}}, \bibinfo {author} {\bibfnamefont
  {K.}~\bibnamefont {Freudenberg}}, \bibinfo {author} {\bibfnamefont
  {P.}~\bibnamefont {Fogarty}}, \bibinfo {author} {\bibfnamefont
  {T.}~\bibnamefont {Hargrove}}, \bibinfo {author} {\bibfnamefont
  {P.}~\bibnamefont {Heitzenroeder}}, \bibinfo {author} {\bibfnamefont
  {G.}~\bibnamefont {Lovett}}, \bibinfo {author} {\bibfnamefont
  {P.}~\bibnamefont {Miller}}, \bibinfo {author} {\bibfnamefont
  {R.}~\bibnamefont {Myatt}}, \bibinfo {author} {\bibfnamefont
  {B.}~\bibnamefont {Nelson}}, \bibinfo {author} {\bibfnamefont
  {W.}~\bibnamefont {Reiersen}}, \ and\ \bibinfo {author} {\bibfnamefont
  {D.}~\bibnamefont {Strickler}},\ }\href {\doibase
  10.1016/j.fusengdes.2005.06.254} {\bibfield  {journal} {\bibinfo  {journal}
  {Fusion Engineering and Design}\ }\textbf {\bibinfo {volume} {75-79}},\
  \bibinfo {pages} {71} (\bibinfo {year} {2005})}\BibitemShut {NoStop}%
\end{thebibliography}%
\end{document}